\documentclass[aps,twocolumn,showlabels,showrefs,amsmath,amssymb,pre,superscriptaddress, floatfix, colors]{revtex4-1}

\usepackage{graphicx}
\usepackage{dcolumn}
\usepackage{bm}
\usepackage{amssymb}
\usepackage{mathtools}
\usepackage{multirow}
\usepackage{bbold}

\newcommand{\1}{\begin{equation}}
\newcommand{\2}{\end{equation}}
\newcommand{\ea}{\begin{eqnarray}} 
\newcommand{\ee}{\end{eqnarray}}

\newcommand{\nablar}{\mathbf{\nabla}_{{\mathbf{r}}}}
\newcommand{\intdr}{\int \! \mathrm{d}\mathbf r}

\begin{document}
\title{Dynamical density functional theory for "dry" and "wet" active matter}
\date{\today}

\pacs{..}

 \author{Hartmut L\"owen}
 \affiliation{Institut f\"{u}r Theoretische Physik II: Weiche Materie, Heinrich-Heine-Universit\"{a}t D\"{u}sseldorf, D-40225 D\"{u}sseldorf, Germany}


\begin{abstract}
In the last 50 years, equilibrium density functional theory (DFT) has been proven to be a powerful, versatile and predictive
approach for the statics and structure of  classical particles. This theory can be extended to the nonequilibrium
dynamics of completely overdamped Brownian colloidal particles towards so-called dynamical density functional theory (DDFT).
The success of DDFT makes it a promising candidate for a first-principle description of active matter. In this lecture,  
we shall first recapitulate classical DDFT for passive colloidal particles typically described by Smoluchowski equation.
After a basic derivation of DDFT from the Smoluchowski equation, we discuss orientational degrees of freedom 
and the effect of hydrodynamic 
interactions for passive particles. This brings us into an ideal position to generalize DDFT towards active matter. 
In particular we 
distinguish between "dry active matter" which is composed of self-propelled particles 
that contain no hydrodynamic flow effects of a surrounding solvent
and "wet active matter" where the hydrodynamic flow fields generated by the microswimmers are taken into account.
For the latter, DDFT is a tool
which unifies thermal fluctuations, direct particle interactions, external driving fields and  hydrodynamic 
effects arising from internal self-propulsion discriminating between "pushers" and "pullers".
 A number of recent applications is discussed including transient clustering of self-propelled rods and
 the spontaneous formation of a hydrodynamic pump
in confined microswimmers.
\end{abstract}
\maketitle

\section{Introduction}

Density functional theory (DFT) relies on the fact that there is a functional of
the one-particle density which gives access to the equilibrium thermodynamics when it is minimized with respect to this density.
This important theory can be both applied to quantum-mechanical electrons and to classical systems.

In this lecture  we shall consider nonequilibrium situations 
for completely overdamped Brownian dynamics of colloids. A dynamical version of 
DFT, the so-called dynamical density functional theory (DDFT), is available and makes 
dynamical predictions which are in good agreement with computer simulations. Here we shall derive
DDFT for Brownian dynamics in a tutorial way from the microscopic Smoluchowski equation.
The theory will then be generalized towards hydrodynamic interactions between the particles 
and to orientational degrees of freedom describing e.g. rod-like colloids. 
This poises us into an ideal position to generalize the DDFT towards active matter systems. For many
interacting active Brownian particles without any hydrodynamic interactions ("dry active matter"), we derive the DDFT approach and discuss 
confinement-induced clustering as one example. Finally we develop  a generic model for microswimmers ("wet active matter")
which includes the hydrodynamic flow field and discriminates between "pushers" and "pullers". In this context, DDFT is a tool
which unifies thermal fluctuations, direct particle interactions, external driving fields with  the hydrodynamic 
effects arising from internal self-propulsion. A number of recent examples relevant for microswimmers
 has been explored within DDFT ranging from the formation of a hydrodynamic pump
in confined system to collections of circle swimmers and binary mixtures of pushers and pullers. 
 For parts of this tutorial 
we follow the ideas outlined in Ref.\ \cite{Springer_book}. 
For more technical aspects, we refer to the recent review 
\cite{review_DDFT}. In contrast to Ref.\ \cite{review_DDFT} these lecture notes are
 {\it not} a balanced review, they are rather a biased tutorial, strongly 
biased with respect to recent works published by the author.


\section{Density Functional Theory (DFT) in equilibrium}

\subsection{Basics}

We shall consider density functional theory (DFT) here for classical systems 
at finite temperature which are interacting via a radially-symmetric  pair-wise potential $v(r)$.
The basic variational principle of
density functional theory establishes the existence of  a unique grand canonical free energy-density-functional $\Omega (T,\mu , [\rho])$, which gets minimal for the equilibrium density $\rho_0(\mathbf{r})$ and then coincides with the real grand canonical free energy, i.e.
\begin{equation}
\left. \frac{\delta \Omega (T, \mu , [\rho])}{\delta \rho(\mathbf{r})} \right|_{\rho(\mathbf{r})=\rho_0(\mathbf{r})} = 0.\label{eq:1.1}
\end{equation}
Here, $T$ is the imposed temperature and $\mu$ the prescribed chemical potential of the system.
However, the functional $\Omega ( T, \mu , [\rho])$ is not known explicitly, in general.
One can split the functional $\Omega ( T, \mu , [\rho])$  as 
\begin{equation}
\Omega(T,\mu,[\rho])=\mathcal{F}(T,[\rho])+\int\limits_{V}d\mathbf{r}\;\rho(\mathbf{r})\left(V_{ext}(\mathbf{r})
-\mu\right)\label{1.2}
\end{equation}
where $\mathcal{F}(T,[\rho])$ is a Helmholtz free energy functional and $V$ denotes the system volume.\\[0,2cm]
The knowledge of the functional $\mathcal{F}(T,[\rho])$ for a given pair potential $v(r)$ 
provides a lot of information (much more than just a bulk equation of state, for instance)
 since it can be applied to any inhomogeneous system in
an external potential $V_{ext}({\mathbf r})$. For example, the second functional derivative taken
in the homogeneous bulk limit is proportional to the direct fluid pair correlation function.\\[0,2cm]

\subsection{Approximations for the density functional}
Let us first recall the exact functional for the ideal gas where the pair interaction $v(r)$ between the particles vanishes, $v(r)=0$. In three spatial dimensions, it reads as
\begin{equation} 
\mathcal{F}(T,[\rho])=\mathcal{F}_{\text{id}}(T,[\rho])= k_B T \int_V \text{d}\mathbf{r}\hspace{4pt}\rho(\mathbf{r})\left[\ln(\rho(\mathbf{r})\Lambda^3)-1\right]
\end{equation}
where $\Lambda$ is the irrelevant thermal wave length and $k_B$ the Boltzmann constant.
In this case, the minimization condition
\begin{equation}
0 = \left.\frac{\delta \Omega}{\delta\rho(\mathbf{r})}\right|_0 = k_B T \ln (\rho(\mathbf{r})\Lambda^3) + V_{\text{ext}}(\mathbf{r}) - \mu
\end{equation}
leads to the generalized barometric law
\begin{equation}
\rho_0(\mathbf{r}) = \frac{1}{\Lambda^3} \exp \left( - \frac{ V_{\text{ext}}(\mathbf{r}) - \mu}{k_B T}\right)\label{eq:density_rho0}
\end{equation}
for the inhomogeneous density. More training in the calculation of functional derivatives will be shifted to the exercises. 
For non-vanishing pair interactions $v(r)$, one can split
\begin{equation} \mathcal{F}(T,[\rho]) =: \mathcal{F}_{\text{id}}(T,[\rho]) + \mathcal{F}_{\text{exc}}(T,[\rho])\end{equation}
which defines a so-called {\it excess\/} free energy density functional ${\cal F}_{exc} (T, [\rho ])$
 which typically needs to be approximated. One important approximation is the {\it mean-field approximation} where
\begin{equation}
{\mathcal F}_\text{exc} (T, [\rho ]) \approx \frac{1}{2} \int \text{d} \mathbf{r} \int \text{d} \mathbf{r}'\, v(| {\vec r} - {\vec r}'| ) \rho( {\vec r})\rho( {\vec r}').
\end{equation}
Other approximations are the perturbative Ramakrishnan-Yussouff (RY) approach or non-perturbative fundamental 
measure theory for steric interactions \cite{mein_review,Roth_review_JPCM_2010}. Similar approximations can be formulated in two spatial dimensions, e.g.\ for hard disks  \cite{Oettel_JCP}.


\section{Classical dynamical density functional theory (DDFT) for passive Brownian particles}	

\subsection{Brownian dynamics and Smoluchowski equation}

DFT can be made time-dependent for passive overdamped 
Brownian particles \cite{review_DDFT} leading to dynamical density functional theory (DDFT).
where the time-dependent density field is the central quantity. It will follow a deterministic diffusion-like equation. 

\subsubsection{Noninteracting Brownian particles}
 For noninteracting particles with an inhomogeneous time-dependent particle density $\rho(\mathbf{r},t)$, 
Fick's law for the current density $\mathbf{j}(\mathbf{r},t)$ states
\begin{equation} \mathbf{j}(\mathbf{r},t)=-D_0\mathbf{\nabla}\rho(\mathbf{r},t)
\end{equation}
where $D_0$ is a phenomenological diffusion coefficient.\par

The continuity equation of particle number conservation
\begin{equation} \frac{\partial\rho(\mathbf{r},t)}{\partial t} + \mathbf{\nabla}\cdot\mathbf{j}(\mathbf{r},t)=0
\end{equation}
then leads to the well-known diffusion equation for $\rho(\mathbf{r},t)$:
\begin{equation}
\frac{\partial \rho(\mathbf{r},t)}{\partial t} = D_0 \Delta \rho(\mathbf{r},t)
\end{equation}
In the presence of an external potential $V_\text{ext}(\mathbf{r},t)$, the force $\mathbf{F}=-\mathbf{\nabla}V_\text{ext}(\mathbf{r},t)$ acts on the particles and will induce a drift velocity $\mathbf{v}_\text{D}$ giving rise to the additional current density 
$\mathbf{j}_\text{D}=\rho\mathbf{v}_\text{D}$ with the drift velocity $ \mathbf{v}_\text{D}=\frac{\mathbf{F}}{\xi}=-\frac{1}{\xi}\mathbf{\nabla}V_\text{ext}(\mathbf{r},t)$. Here,  $\xi$ denotes the friction coefficient (for a sphere of radius $R$ in a viscous solvent of viscosity $\eta$ Stokes law tells us that  $\xi = 6 \pi \eta R$). With the Stokes-Einstein relation   $D_0=\frac{k_BT}{\xi}$ we get
 $\mathbf{j}=-\frac{1}{\xi}(k_BT\mathbf{\nabla}\rho+\rho\mathbf{\nabla}V_\text{ext})$ and the continuity equation yields
\begin{equation}
\frac{\partial\rho(\mathbf{r},t)}{\partial t} = \frac{1}{\xi}(k_BT\Delta\rho(\mathbf{r},t)+\mathbf{\nabla}\cdot (\rho(\mathbf{r},t)\mathbf{\nabla}V_\text{ext}(\mathbf{r},t)))
\end{equation}
which is the \underline{Smoluchowski equation} for non-interacting particles. Note that the external force can even be time-dependent.

\subsubsection{Interacting Brownian particles}
Now we consider $N$ \textbf{interacting} particles at positions $\vec {r}_i$ ($i=1,...,N$). The total potential energy is
\begin{equation}U_\text{tot}(\mathbf{r}^N,t)=\displaystyle\sum_{i=1}^NV_\text{ext} (\mathbf{r}_i,t)+\sum_{\substack{i,j=1\\i<j}}^Nv(|\mathbf{r}_i-\mathbf{r}_j|)
\end{equation}
With $\mathbf{r}^N = \{\mathbf{r}_i \; (i=1, \dots, N)\}$ the generalization of the Smoluchowski equation \cite{Doi_Edwards} for the joint probablity density
$p(\mathbf{r}^N,t)$ reads as 
\begin{equation}\frac{\partial p}{\partial t}=\hat{\mathcal{O}} p=\frac{1}{\xi} \displaystyle\sum_{i=1}^N\mathbf{\nabla}_i\cdot[k_BT\mathbf{\nabla}_i+\mathbf{\nabla}_iU_\text{tot} (\mathbf{r}^N,t)]p
\end{equation}
where the operator $\hat{\mathcal{O}}$ is called Smoluchowski operator.

\subsection{Derivation of DDFT}

\subsubsection{Phenomenological derivation of DDFT}

 The general Fick's law assumes that the particle current density is proportional to the gradient of the chemical potential
\cite{Doi_Edwards} 
 and proportional to the time-dependent density $\rho(\mathbf{r},t)$:
\begin{equation}
\mathbf{j}= \xi \rho(\mathbf{r},t)\mathbf{\nabla}\mu
\end{equation}
 In equilibrium, when the chemical potential is constant, 
there is no such current. We now take a functional derivative with respect to the density in Eq.~(\ref{1.2}) and  obtain in the 
absence of an external potential
\begin{equation}
\left. \frac{\delta {\cal F} (T, [\rho])}{\delta \rho(\mathbf{r})} \right|_{\rho(\mathbf{r})=\rho_0(\mathbf{r})} = \mu\label{DDFT1}
\end{equation}
When combining this with the continuity equation of particle number conservation we get the important DDFT equation
\begin{equation}
\label{eqn:DDFT}
\xi\frac{\partial\rho(\mathbf{r},t)}{\partial t}=\mathbf{\nabla}\rho(\mathbf{r},t) \mathbf{ \nabla} \frac{\delta{\cal F}[\rho]}{\delta\rho(\mathbf{r},t)}
\end{equation}
which is obviously generalized to the presence of an external potential $V_\text{ext}(\mathbf{r},t)$
by replacing ${\cal F}[\rho]$ with $\Omega[\rho]$.
This is a deterministic time evolution equation for $\rho(\mathbf{r},t)$. For an ideal gas, it reduces to the exact
Smoluchowski equation which is
standard diffusion equation, see exercise. For an interacting system, the DDFT equation is approximative.

\subsubsection{Derivation of DDFT from the Smoluchowski equation}

The DDFT equation can be derived from  the Smoluchowski equation \cite{Archer_Evans}
 but one essential additional approximation, the so-called adiabatic approximation, needs to be performed here as well.
In more detail, one integrates out degrees of freedom from the Smoluchowski equation to obtain the following exact equation
\begin{align}
\xi\frac{\partial}{\partial t}\rho(\mathbf{r}_1,t)= {} &k_BT\Delta_1\rho(\mathbf{r}_1,t)+ \mathbf{\nabla}_1(\rho(\mathbf{r}_1,t)\mathbf{\nabla}_1V_\text{ext}(\mathbf{r}_1,t)\nonumber\\
&+\mathbf{\nabla}_1\int\text{d}\mathbf{r}_2\hspace{4pt}\rho^{(2)}(\mathbf{r}_1,\mathbf{r}_2,t) \mathbf{\nabla}_1v(|\mathbf{r}_1-\mathbf{r}_2|)
\end{align}
In equilibrium, necessarily $\frac{\partial\rho(\mathbf{r}_1,t)}{\partial t}=0$ which implies
\begin{align}
0&= \mathbf{\nabla}\left(k_BT\mathbf{\nabla}\rho(\mathbf{r})+\rho(\mathbf{r})\mathbf{\nabla}V_\text{ext} (\mathbf{r})\right. \\
&\left.+\int\text{d} \mathbf{r}^{'}\hspace{4pt}\rho^{(2)}(\mathbf{r},\mathbf{r}^{'})\mathbf{\nabla}v(|\mathbf{r}- \mathbf{r}^{'}|)\right)
\end{align}
menaing that the divergence of a current density must vanish. The current density itself is imposed to vanish for $r\to\infty$ in equilibrium and  thus the curent density is identical to zero everywhere. Therefore
\begin{equation}
 0=k_BT\mathbf{\nabla}\rho(\mathbf{r})+\rho(\mathbf{r})\mathbf{\nabla}V_\text{ext}(\mathbf{r}) + \int\text{d}\mathbf{r}^{'}\hspace{4pt}\rho^{(2)}(\mathbf{r},\mathbf{r}^{'})\mathbf{\nabla}v(|\mathbf{r}-\mathbf{r}^{'}|)
\end{equation}
which is also known as Yvon-Born-Green-relation (YBG). Here, $\rho^{(2)}(\mathbf{r},\mathbf{r}^{'})$ is the two-body joint probability density in nonequilibrium. We now take a gradient of the density functional derivative of  Eq.~(\ref{1.2})
and combine it with YBG-relation. Then we obtain
\begin{equation}\label{eq:adiabatic_approx}
\int\text{d}\mathbf{r}'\hspace{4pt} \rho^{(2)}(\mathbf{r},\mathbf{r}')\mathbf{\nabla}V(|\mathbf{r}-\mathbf{r}'|)=\rho(\mathbf{r})\mathbf{\nabla}\frac{\delta \mathcal{F}_\text{exc}[\rho]}{\delta \rho(\mathbf{r})}
\end{equation}
We postulate that this argument holds also in nonequilibrium. In doing so, non-equilibrium correlations are approximated by equilibrium ones at the same $\rho(\mathbf{r},t)$ (identified via a suitable time-independent $V_\text{ext}(\mathbf{r})$ in equilibrium).
Equivalently, one can say that it is postulated that pair correlations decay much faster to their equilibrium one than the one-body density \cite{JCP_2009}. 
 This results after all in the DDFT equation:
\begin{equation}
\label{eqn:DDFT}
\xi\frac{\partial\rho(\mathbf{r},t)}{\partial t}=\mathbf{\nabla}\rho(\mathbf{r},t) \mathbf{ \nabla} \frac{\delta{\Omega}[\rho]}{\delta\rho(\mathbf{r},t)}
\end{equation}
For further alternate derivation of the DDFT equation, see \cite{Marconi_1,Marconi_2,JCP_2009}.

\section{Polar particles}


\subsection{DFT of polar particles}									

We now consider polar particles which possess an additional rotational degree of freedom in the two-dimensional plane which can be described by a simple angle $\phi$ or a unit vector 
\begin{equation}
\hat{\mathbf{n}}=(\cos \phi, \sin \phi )
\end{equation}
 relative to a prescribed axis.
Having applications to swimmers on a substrate in mind, we consider motion in two-dimensions only.
Equilibrium DFT can readily be extended to polar particles.
A configuration of $N$ particles is now fully specified by the set of positions of the center of masses and the corresponding orientations 
$\{\mathbf{r}_i,\hat{\mathbf{n}}_i,i=1,\ldots,N\}$. Pairwise interactions are described by a pair-potential
$v(\mathbf{r}_i-\mathbf{r}_j,\hat{\mathbf{n}}_i,\hat{\mathbf{n}}_j)$ that depends on the difference vector $\mathbf{r}_i-\mathbf{r}_j$ between the centers of the
particle $i$ and another particle $j$ plus their two orientations $\hat{\mathbf{n}}_i$ and $\hat{\mathbf n}_j$.
In the general context of active matter, if this function {\it only} depends on $\mathbf{r}_i-\mathbf{r}_j$, 
the interactions are called {\it non-aligning}. An example are 
spherical self-propelled Janus particles which do not change their orientation when bouncing into each other. 
If it is energetically favorable to have parallel
orientations, the interactions are called  {\it aligning}. 
In the rare case that neighbouring particle tend to stay anti-parallel these interactions are called 
 {\it anti-aligning}. 
Clearly the external potential $V_{\text{ext}} (\mathbf{r},\hat{\mathbf{n}},t)$ can also depend on the particle orientation.

As in the case of spherical particles, DFT establishes the existence of a functional
of the  one-particle density  $\rho(\mathbf{r},\hat{\mathbf{n}})$ which gets minimal in equilibrium
\begin{equation}
\left.\frac{\delta\Omega(T,\mu,[\rho])}{\delta\rho(\mathbf{r},\hat{\mathbf{n}})}\right|_{ \rho=\rho(\mathbf{r},\hat{\mathbf{n}})}=0
\end{equation}
Again, the functional can be decomposed as follows
\begin{align}\label{eqn:decompose}
&\Omega(T,\mu,[\rho])= {} k_BT\int\text{d}\mathbf{r}\int_0^{2\pi}\text{d}\phi \hspace{4pt}\rho(\mathbf{r},\hat{\mathbf{n}}) [\ln(\Lambda^2\rho(\mathbf{r},\hat{\mathbf{n}}))-1]\nonumber\\
&+\int\text{d}\mathbf{r}\int_0^{2\pi}\text{d}\phi(V_\text{ext}(\mathbf{r},\hat{\mathbf{n}})-\mu) \rho(\mathbf{r},\hat{\mathbf{n}})+\mathcal{F}_\text{exc}(T,[\rho])
\end{align}
The first term on the right hand side of equation (\ref{eqn:decompose}) is the functional $\mathcal{F}_\text{id}[\rho^{(1)}]$ for ideal rotators. The excess part $\mathcal{F}_\text{exc}(T,[\rho^{(1)}])$ is in general unknown and requires approximative treatments.
Again nonperturbative fundamental measure theory for hard cylinders is available \cite{Wittmann}.


\subsection{DDFT for polar particles}											

The Smoluchowski equation for the joint probability density 
$p(\mathbf{r}_1,\cdots,\mathbf{r}_N;\hat{\mathbf{n}}_1,\cdots,\hat{\mathbf{n}}_N,t)=p(\mathbf{r}^N, \hat{\mathbf{n}}^N,t)$ is 
\begin{equation}
\frac{\partial p}{\partial t} = \hat{\mathcal{O}}_Sp
\end{equation}
with 
 the Smoluchowski operator 
\begin{align}
\hat{\mathcal{O}}_S = {} &\sum\limits_{i=1}^N\left[ \nabla_{\mathbf{r}_i}\cdot\bar{\bar{D}}(\hat{\mathbf{n}}_i)\cdot\left(\mathbf{\nabla}_{\mathbf{r}_i} +\frac{1}{k_BT}\mathbf{\nabla}_{\mathbf{r}_i}U(\mathbf{r}^N,\hat{\mathbf{n}}^N,t)\right)\right.\nonumber\\
&\left.+D_r\hat{\mathbf{R}}_i\cdot\left(\hat{\mathbf{R}}_i+\frac{1}{k_BT}\hat{\mathbf{R}}_iU(\mathbf{r}^N,\hat{\mathbf{n}}^N,t)\right)\right]
\end{align}
where $U(\mathbf{r}^N,\hat{\mathbf{n}}^N,t)$ is the total potential energy. Here the rotation operator $\hat{\mathbf{R}}_i$ 
is defined as $\hat{\mathbf{R}}_i=\partial /\partial \phi$  and the anisotropic translational diffusion tensor is given by
\begin{equation}
\bar{\bar{D}}(\hat{\mathbf{n}}_i)=D^\shortparallel\hat{\mathbf{n}}_i\otimes \hat{\mathbf{n}}_i+D^\perp (\mathbf{1}-\hat{\mathbf{n}}_i\otimes\hat{\mathbf{n}}_i)
\end{equation}
The two diffusion constants $D^\shortparallel$ and $D^\perp$, parallel and perpendicular to the orientations reflect the fact that the translational diffusion is anisotropic. The quantity $D_r$ is called rotational diffusion constant and sets the Brownian dynamics of the orientations.

Integrating the Smoluchowski equation yields the following DDFT equation for the time-dependent 
$\rho(\mathbf{r},\phi,t)$ \cite{Rex_Wensink_2007}
\begin{align}\label{eq:115_2d}
\frac{\partial \rho(\mathbf{r},\phi,t)}{\partial t} = {} & \nablar \cdot \bar{\bar{D}}(\phi)\cdot\left[ \rho(\mathbf{r},\phi,t)\nablar \frac{\delta\Omega[\rho(\mathbf{r},\phi,t)]}{\delta \rho(\mathbf{r},\phi,t)}\right]\nonumber\\
&+ D_r\frac{\partial }{\partial \phi} \left[\rho(\mathbf{r},\phi,t) \frac{\partial }{\partial \phi} \frac{\delta\Omega[\rho(\mathbf{r},\phi,t)]}{\delta \rho(\mathbf{r},\phi,t)}\right]
\end{align}

\section{Dynamical density functional theory for active Brownian particles ("dry" active matter)}


A simple classification of active matter can be done into "dry active matter" where solvent flow does not play any role and "wet active matter"
where hydrodynamic effects are important. In this chapter we shall study the simpler case of dry active matter first and treat wet active matter in the next chapter. 
Ignoring hydrodynamic interactions, these swimmers can simply be
modeled by  polar particles which are driven by a constant internal effective force
along their orientations \cite{JPCM_2015}; this force corresponds to an effective drift
velocity and mimics the actual propulsion mechanism.
On top of the intrinsic propulsion, the particles feel Brownian noise of the solvent. The corresponding
motion is intrinsically a nonequilibrium one and even the dynamics of a
single Brownian swimmer was solved only in this century \cite{Golestanian,Teeffelen_PRE_2008,Hagen_JPCM}.

For dry active matter, the many-body Smoluchowski equation now reads 
\begin{equation}
\frac{\partial p}{\partial t} = \hat{\mathcal{O}}_ap 
\end{equation}
with the "active" Smoluchowski operator
\begin{align}
\hat{\mathcal{O}}_a = \hat{\mathcal{O}}_S + {} &\sum\limits_{i=1}^N\left[ \nabla_{\mathbf{r}_i}\cdot\bar{\bar{D}}(\hat{\mathbf{n}}_i)\cdot
\left(   \frac{1}{k_BT} v_0 \hat{\mathbf{n}}_i \right) \right]
\end{align}
The active part involves a particle current along the particle orientation with a strength proportional to $v_0$ which
is the self-propulsion velocity a single particle assumes. This Smoluchowski equation is stochastically equivalent to 
the Langevin equations of active Brownian motion \cite{our_RMP}.
For ideal particles ($U(\mathbf{r}^N,\hat{\mathbf{n}}^N,t)=0$), the active Smoluchowski equation has been the starting point
to calculate  the intermediate scattering function
of an active Brownian particle \cite{Kurzthaler_2016}.

 DDFT for dry active matter can be derived using the same adiabatic
approximation \eqref{eq:adiabatic_approx} as in the passive case.
The resulting equation of motion
for the one-particle density $\rho(\mathbf{r},\phi,t)$ then reads  \cite{Wensink_2008}:
\begin{align}\label{eq:dry_1}
&k_BT\frac{\partial \rho(\mathbf{r},\phi,t)}{\partial t} =  \nablar \cdot \bar{\bar{D}}(\phi)\cdot\left[k_BT v_0 \hat{\mathbf{n}}\rho(\mathbf{r},\phi,t)\frac{\vphantom{h}}{\vphantom{h}}\right. \nonumber\\
&\left. + \rho(\mathbf{r},\phi,t)\nablar \frac{\delta\Omega[\rho(\mathbf{r},\phi,t)]}{\delta \rho(\mathbf{r},\phi,t)} \right] \nonumber \\
&+ D_r \frac{\partial }{\partial \phi} \left[\rho(\mathbf{r},\phi,t) \frac{\partial }{\partial \phi} \frac{\delta\Omega[\rho(\mathbf{r},\phi,t)]}{\delta \rho(\mathbf{r},\phi,t)}\right]
\end{align}
For a non-interacting system, it is important to note here that this equation 
is exact under self-propulsion and any external forces, see exercises. It is therefore an ideal starting point to study
a single active Brownian particle under gravity \cite{Enculescu_Stark,Kuemmel,Herrmann_Schmidt,Mazza}. As a result,
 polar order was discovered 
for an ideal gas of sedimentating particles in the steady state even if the particle are not bottom-heavy.

As an application, for active particles in a channel with aligning interactions, the
time-dependent density profiles were found to be in agreement with Brownian dynamics
computer simulations \cite{Wensink_2008}. In Ref.\ \cite{Wensink_2008}, a crude mean-field
Onsager-like density functional approximation
\cite{Onsager}
was  used.  Qualitatively, a transient
formation of hedgehog-like clusters near the channel boundaries was found in simulations and
reproduced by the DDFT.

Finally we remark that  DDFT was generalized towards three spatial dimensions for swimmers
of arbitrary shape with complicated friction tensors \cite{Witt_Mol_Phys}. 
Moreover, superadiabatic DDFT which goes beyond the adiabatic approximation 
has been applied to active Brownian systems with  non-aligning interactions \cite{Krinninger_2016,Krinninger_2019}.
A special application was performed for  motility-induced phase separation of active particles \cite{Krinninger_2016,Hermann_2019}.

\section{Dynamical density functional theory for microswimmers ("wet" active matter)}

The most general DDFT framework for microswimmers can be found in Ref.\ \cite{Menzel_JChemPhys_2016} which we closely follow here. 
This approach includes
simultaneously thermal fluctuations, external forces, interparticle interactions by body forces and hydrodynamic interactions 
as well as self-propulsion effects.
In principle it includes all previous cases in special limit of vanishing self-propulsion ("limit of "passive particles") 
and dry active matter (limit of neglected hydrodynamic interactions).

\subsection{The swimmer model}

\begin{figure}
\includegraphics[width=7.3cm]{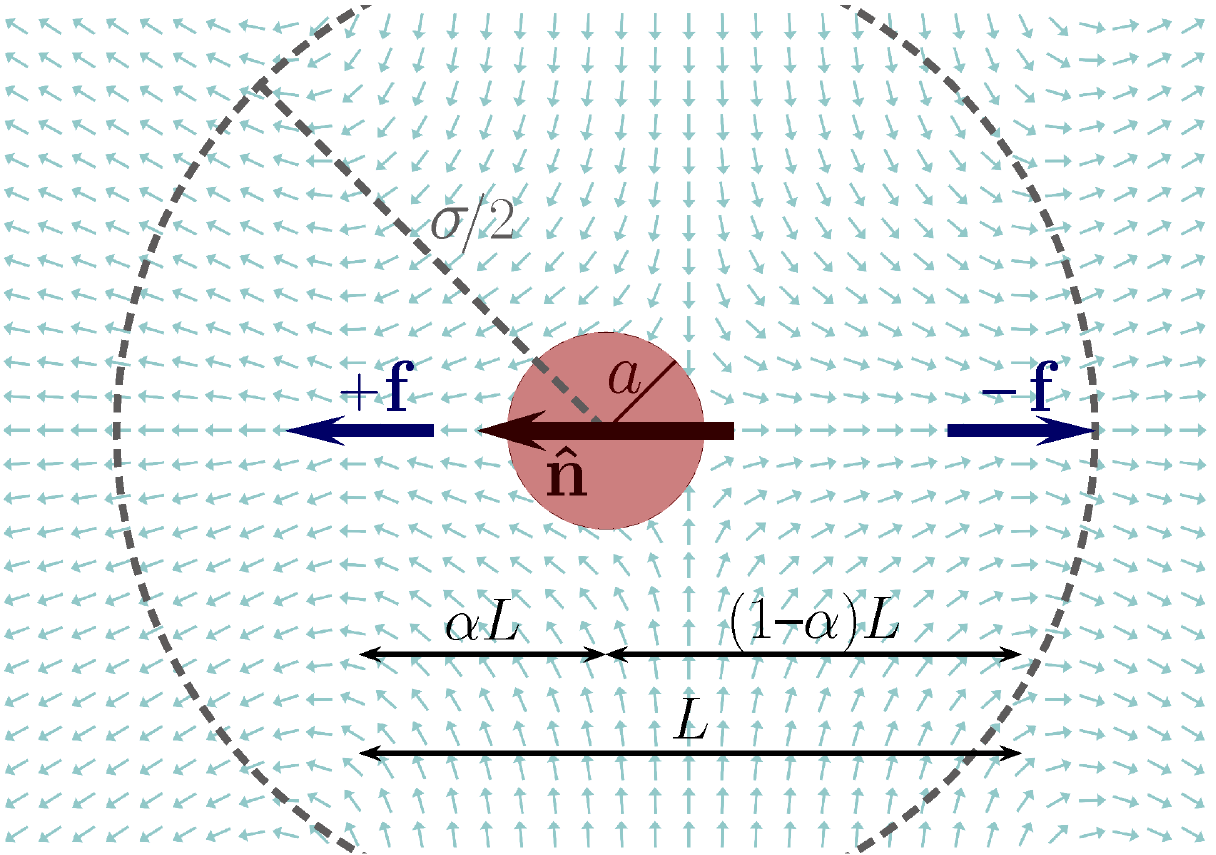}
\caption{
Individual model microswimmer. The spherical swimmer body of hydrodynamic radius $a$ is subjected to hydrodynamic drag. Two active point-like force centers exert active forces $+{\bf f}$ and $-{\bf f}$ onto the surrounding fluid. This results in a self-induced fluid flow indicated by small light arrows. $L$ is the distance between the two force centers. The whole set-up is axially symmetric with respect to the axis $\mathbf{\hat n}$. If the swimmer body is shifted along $\mathbf{\hat n}$ out of the geometric center, leading to distances $\alpha L$ and $(1-\alpha)L$ to the two force centers, it feels a net self-induced hydrodynamic drag. The microswimmer then self-propels. In the depicted state (pusher), fluid is pushed outward. Upon inversion of the two forces, fluid is pulled inward (puller). We consider soft isotropic steric interactions between the swimmer bodies of typical interaction range $\sigma$, implying an effective steric swimmer radius of $\sigma/2$. From Ref.\ \cite{Menzel_JChemPhys_2016}. }
\label{fig_microswimmer}
\end{figure}

To derive the DDFT, we consider a dilute suspension of $N$ identical self-propelled microswimmers at low Reynolds number \cite{purcell1977life} in two dimensions
in an unbounded three dimensional fluid. 
Following Ref. \cite{Menzel_JChemPhys_2016}, the self-propulsion of a microswimmer is 
concatenated to self-induced fluid flows in the surrounding medium. 
This then represents a major source of hydrodynamic interaction between different swimmers. 
To proceed we consider a minimal model for an individual microswimmer as depicted in Fig.~\ref{fig_microswimmer}. 
Each microswimmer consists of a spherical body of hydrodynamic radius $a$. The swimmer body is subjected to hydrodynamic drag with respect to surrounding fluid flows including self-convection. The latter is generated by two active force centers which are located at a distance $L$ from each other, see Fig.~\ref{fig_microswimmer}, and exert two antiparallel forces $+\mathbf f$ and $-\mathbf f$, respectively, onto the surrounding fluid and set it into motion. Summing up the two forces, we find that the microswimmer exerts a vanishing net force onto the fluid. Moreover, since $\mathbf f\|{\mathbf{\hat n}}$, there is no net active torque \cite{fily2012cooperative}. Self-propulsion is now achieved by shifting the swimmer body along $\mathbf{\hat n}$ out of the geometric center. We introduce a parameter $\alpha$ to quantify this shift, see Fig.~\ref{fig_microswimmer}. The distances between the body center and the force centers are now $\alpha L$ and $(1-\alpha)L$, respectively. We confine $\alpha$ to the interval $]0,0.5]$. For $\alpha=0.5$, no net self-induced motion occurs by symmetry. 
For $\alpha\neq0.5$, the swimmer body feels a net self-induced fluid flow due to the proximity to one of the two force centers. Due to the resulting self-induced hydrodynamic drag on the swimmer body, the swimmer self-propels. It is important to note that the two force centers are completely fixed or attached to the particle center. So they propel the particle and are moving with the particle at the same time. In the depicted state of outward oriented forces, the swimmer pushes the fluid outward and is called a {\it pusher\/} \cite{baskaran2009statistical}. Inverting the forces, the swimmer pulls fluid inward and is termed a {\it puller\/} \cite{baskaran2009statistical}.

We now consider an assembly of $N$ interacting identical self-propelled model microswimmers, suspended in a viscous, incompressible fluid at low Reynolds number \cite{purcell1977life}. The flow profile within the system then follows Stokes' equation \cite {happel2012low},
\begin{eqnarray}
\eta\nabla^2\mathbf v(\mathbf r,t)+\nabla p(\mathbf r,t)=\sum_{i=1}^N\mathbf f_i(\mathbf r_i,\mathbf{\hat n}_i,t). 
\label{Stokes}
\end{eqnarray}
Here, $t$ denotes time and $\mathbf r$ any spatial position in the suspension, while $\mathbf v(\mathbf r,t)$ gives the
 corresponding fluid flow velocity field. $\eta$ is the viscosity of the fluid and $p(\mathbf r,t)$ is the pressure field. On the right-hand side, 
$\mathbf f_i$ denotes the total force density field exerted by the $i$th microswimmer onto the fluid. $\mathbf r_i$ and $\mathbf{\hat n}_i$ mark the 
current position and orientation of the $i$th swimmer at time $t$, respectively. This each microswimmer contributes to the overall fluid flow in the system by the force density it exerts on the fluid.  In this way, each swimmer can transport itself via active self-propulsion since the point force centers are firmly attached to the swimmer body.   Moreover, all swimmers hydrodynamically interact with each other via their induced flow fields. 

The linearity of Eq.~(\ref{Stokes}) and the incompressibility of the fluid, i.e. $\nabla \cdot \mathbf v (\mathbf r,t)=0$, 
implies a linear relation between velocities (angular velocities) and forces (torques).

We denote by $\mathbf F_j$ and $\mathbf T_j$ the forces and torques, respectively, acting directly on the swimmer bodies ($j=1,...,N$), except for frictional forces and  torques resulting from the surrounding fluid. 
The non-hydrodynamic body forces and torques may for example result from external potentials or steric interactions and will be specified below. From them, in the passive case, i.e.\ for $\mathbf f=\mathbf {0}$, the instantly resulting velocity  $\mathbf v_i$ and angular velocity $\bm{\omega}_i$ of the $i$th swimmer body follows as 
\begin{equation}
     \begin{bmatrix}
         \mathbf v_i \\
          \bm{\omega}_i
     \end{bmatrix}
     =
    \sum_{j=1}^N
    {\bf {M}}_{ij}\cdot\begin{bmatrix}
        \mathbf F_j \\
        \mathbf T_j
    \end{bmatrix}
    =
    \sum_{j=1}^N
    \begin{bmatrix}
    \bm{\mu}^{tt}_{ij} & \bm{\mu}^{tr}_{ij}\\
    \bm{\mu}^{rt}_{ij} & \bm{\mu}^{rr}_{ij}    
    \end{bmatrix}
    \cdot
    \begin{bmatrix}
        \mathbf F_j \\
        \mathbf T_j
    \end{bmatrix}.  
    \label{Stokes1}  
\end{equation} 
Here ${\bf M}_{ij}$ are the mobility matrices, the components of which ($\bm{\mu}^{tt}_{ij}$, $\bm{\mu}^{tr}_{ij}$, $\bm{\mu}^{rt}_{ij}$, $\bm{\mu}^{rr}_{ij}$) likewise form matrices.  
They describe hydrodynamic translation--translation, translation--rotation, rotation--translation, and rotation--rotation coupling, respectively.

The mobility matrices can approximately be calculated as
\begin{eqnarray}
\bm{\mu} ^{tt}_{ii}&=&\mu^t{\bf 1},
\quad\bm{\mu}_{ii}^{rr}=\mu^r{\bf 1},
\quad\bm{\mu}_{ii}^{tr}=\bm{\mu}_{ii}^{rt}=\mathbf{0}\quad
\label{Stokes2}
\end{eqnarray} 
for entries $i=j$ (no summation over $i$ in these expressions) and
\begin{eqnarray}
\bm{\mu}_{ij}^{tt}&=&\mu^t\bigg(\frac{3a}{4r_{ij}}\Big({\bf {1}}+{{\mathbf{\hat r}_{ij}\mathbf{\hat r}_{ij}}}\Big)  \nonumber\\ 
&&{}+\frac{1}{2}\Big(\frac{a}{r_{ij}}\Big)^3\Big({\bf {1}}-3{{\mathbf{\hat r}_{ij}\mathbf{\hat r}_{ij}}}\Big)\bigg), 
\label{mu_tt} \\ 
\bm{\mu}_{ij}^{rr}&=&{}-\mu^r\frac{1}{2}\left(\frac{a}{r_{ij}}\right)^3\left({\bf {1}}-3{{\mathbf{\hat r}_{ij}\mathbf{\hat r}_{ij}}}\right), \\ 
\bm{\mu}_{ij}^{tr}&=&\bm{\mu}_{ij}^{rt}=\mu^r\left(\frac{a}{r_{ij}}\right)^3{ {\mathbf r_{ij}}}\times, 
\label{mu_tr}
\end{eqnarray} 
for entries $i\neq j$. Here, we have introduced the abbreviations
\begin{equation}\label{abbr}
\mu^t=\frac{1}{6\pi\eta a}, \qquad \mu^r=\frac{1}{8\pi\eta a^3}.
\end{equation}   
 Because of the linearity of Eq.~(\ref{Stokes}), the effect of the active forces
can be added to the swimmer velocities and angular velocities on the right-hand side of Eq.~(\ref{Stokes1}). 


\begin{equation}
\begin{aligned}
\left[\begin{array}{c}
\mathbf{v}_{i} \\
\omega_{i}
\end{array}\right]=& \sum_{j=1}^{N}\left(\left[\begin{array}{cc}
\boldsymbol{\mu}_{i j}^{\mathrm{tt}} & \boldsymbol{\mu}_{i j}^{\mathrm{tr}} \\
\boldsymbol{\mu}_{i j}^{\mathrm{rt}} & \boldsymbol{\mu}_{i j}^{\mathrm{rr}}
\end{array}\right] \cdot\left[\begin{array}{c}
\mathbf{F}_{j} \\
\mathbf{T}_{j}
\end{array}\right]+\left[\begin{array}{cc}
\mathbf{\Lambda}_{i j}^{\mathrm{tt}} & \mathbf{0} \\
\boldsymbol{\Lambda}_{i j}^{\mathrm{rt}} & \boldsymbol{0}
\end{array}\right] \cdot\left[\begin{array}{c}
f \hat{\mathbf{n}}_{j} \\
\mathbf{0}
\end{array}\right]\right)
\end{aligned}
\end{equation}

Note that there are no active torques here, i.e.\ we are considering a linear swimmer here. For circle swimmers,
 a constant torque must be included to describe the circling. Moreover, $\mathbf{\Lambda}_{i j}^{\mathrm{tt}}$, $\mathbf{\Lambda}_{i j}^{\mathrm{rt}}$, summarize effect of both $+f\hat{\mathbf n}_j$ and $-f\hat{\mathbf n}_j$ such that the total swimmer is force-free. In detail, 
\begin{equation*}
\mathbf{\Lambda}_{i j}^{\mathrm{tt}} = \boldsymbol{\mu}_{i j}^{\mathrm{tt+}} - \boldsymbol{\mu}_{i j}^{\mathrm{tt-}}, \mathbf{\Lambda}_{i j}^{\mathrm{rt}} = \boldsymbol{\mu}_{i j}^{\mathrm{rt+}} - \boldsymbol{\mu}_{i j}^{\mathrm{rt-}}
\end{equation*}
For $i=j$, the term  $\mathbf{\Lambda}^{\mathrm{tt}}_{ii}$ contains the self-propulsion of the particles.

\subsection{Derivation of the DDFT for microswimmers}

We now aim at a statistical description for full joint probability density $P=P(\mathbf r^N, \hat{\mathbf{n}}^N, t)$ and start from the  dynamical Smoluchowski equation
\begin{equation}
\frac{\partial P}{\partial t}=-\sum_{i=1}^{N}\left\{\nabla_{\mathbf{r}_{i}} \cdot\left(\mathbf{v}_{i} P\right)+\left(\hat{\mathbf{n}}_{i} \times \nabla_{\hat{\mathbf{n}}_{i}}\right) \cdot\left(\omega_{i} P\right)\right\}
\end{equation}
Integrating out all degrees of freedom except for one swimmer, we get the following
 exact relation for the dynamics of the 
swimmer one-body density 
\begin{widetext}
\begin{equation}
\frac{\partial \rho^{(1)}(\mathbf{X}, t)}{\partial t}=-\nabla_{\mathbf{r}} \cdot\left(\mathcal{J}^{\mathrm{tt}}+\mathcal{J}^{\mathrm{tr}}+\mathcal{J}^{\mathrm{ta}}\right)-\left(\hat{\mathbf{n}} \times \nabla_{\hat{\mathbf{n}}}\right) \cdot\left(\mathcal{J}^{\mathrm{rt}}+\mathcal{J}^{\mathrm{rr}}+\mathcal{J}^{\mathrm{ra}}\right)
\end{equation}
where $\mathbf{X} =  (\bf r, \hat{\bf n})$ is a compact notation for both
translational and orientational degrees of freedom. The six current densities are given by


\begin{equation}
\begin{aligned}
\mathcal{J}^{\mathrm{tt}}=&-\mu^{\mathrm{t}}\left(k_{\mathrm{B}} T \nabla_{\mathbf{r}} \rho^{(1)}(\mathbf{X}, t)+\rho^{(1)}(\mathbf{X}, t) \nabla_{\mathbf{r}} V_{\mathrm{ext}}(\mathbf{r})+\int \mathrm{d} \mathbf{X}^{\prime} \rho^{(2)}\left(\mathbf{X}, \mathbf{X}^{\prime}, t\right) \nabla_{\mathbf{r}} v\left(|\mathbf{r}-\mathbf{r}^{\prime}|\right)\right) \\
&-\int \mathrm{d} \mathbf{X}^{\prime} \boldsymbol{\mu}_{\mathbf{r}, \mathbf{r}^{\prime}}^{\mathrm{tt}} \cdot\left(k_{\mathrm{B}} T \nabla_{\mathbf{r}^{\prime}} \rho^{(2)}\left(\mathbf{X}, \mathbf{X}^{\prime}, t\right)+\rho^{(2)}\left(\mathbf{X}, \mathbf{X}^{\prime}, t\right) \nabla_{\mathbf{r}^{\prime}} V_{\mathrm{ext}}\left(\mathbf{r}^{\prime}\right)\right.\\
&\left.+\rho^{(2)}\left(\mathbf{X}, \mathbf{X}^{\prime}, t\right) \nabla_{\mathbf{r}^{\prime}} v\left(|\mathbf{r}-\mathbf{r}^{\prime}|\right)+\int \mathrm{d} \mathbf{X}^{\prime \prime} \rho^{(3)}\left(\mathbf{X}, \mathbf{X}^{\prime}, \mathbf{X}^{\prime \prime}, t\right) \nabla_{\mathbf{r}^{\prime}} v\left(|\mathbf{r}^{\prime}-\mathbf{r}^{\prime\prime}|\right)\right) \\
\mathcal{J}^{\mathrm{tr}}=&-\int \mathrm{d} \mathbf{X}^{\prime} k_{\mathrm{B}} T \boldsymbol{\mu}_{\mathbf{r}, \mathbf{r}^{\prime}}^{\mathrm{tr}}\left(\hat{\mathbf{n}}^{\prime} \times \nabla_{\hat{\mathbf{n}}^{\prime}}\right) \rho^{(2)}\left(\mathbf{X}, \mathbf{X}^{\prime}, t\right) \\
\mathcal{J}^{\mathrm{ta}}=& f\left(\mathbf{\Lambda}_{\mathbf{r}, \mathbf{r}}^{\mathrm{tt}} \cdot \hat{\mathbf{n}} \rho^{(1)}(\mathbf{X}, t)+\int \mathrm{d} \mathbf{X}^{\prime} \mathbf{\Lambda}_{\mathbf{r}, \mathbf{X}^{\prime}}^{\mathrm{tt}} \cdot \hat{\mathbf{n}}^{\prime} \rho^{(2)}\left(\mathbf{X}, \mathbf{X}^{\prime}, t\right)\right) \\
\mathcal{J}^{\mathrm{rt}}=&-\int \mathrm{d} \mathbf{X}^{\prime} \boldsymbol{\mu}_{\mathbf{r}, \mathbf{r}^{\prime}}^{\mathrm{rt}}\left(k_{\mathrm{B}} T \nabla_{\mathbf{r}^{\prime}} \rho^{(2)}\left(\mathbf{X}, \mathbf{X}^{\prime}, t\right)+\rho^{(2)}\left(\mathbf{X}, \mathbf{X}^{\prime}, t\right) \nabla_{\mathbf{r}^{\prime}} V_{\mathrm{ext}}\left(\mathbf{r}^{\prime}\right)\right.\\
&\left.+\rho^{(2)}\left(\mathbf{X}, \mathbf{X}^{\prime}, t\right) \nabla_{\mathbf{r}^{\prime}} v\left(|\mathbf{r}-\mathbf{r}^{\prime}|\right)+\int \mathrm{d} \mathbf{X}^{\prime \prime} \rho^{(3)}\left(\mathbf{X}, \mathbf{X}^{\prime}, \mathbf{X}^{\prime \prime}, t\right) \nabla_{\mathbf{r}^{\prime}} v\left(|\mathbf{r}^{\prime}-\mathbf{r}^{\prime\prime}|\right)\right) \\
\mathcal{J}^{\mathrm{rr}} =&-k_{\mathrm{B}} T \mu^{\mathrm{r}} \hat{\mathbf{n}} \times \nabla_{\mathrm{n}} \rho^{(1)}(\mathbf{X}, t)-\int \mathrm{d} \mathbf{X}^{\prime} k_{\mathrm{B}} T \boldsymbol{\mu}_{\mathbf{r}, \mathbf{r}^{\prime}}^{\mathrm{rr}} \cdot\left(\hat{\mathbf{n}}^{\prime} \times \nabla_{\mathbf{n}^{\prime}}\right) \rho^{(2)}\left(\mathbf{X}, \mathbf{X}^{\prime}, t\right), \\
\mathcal{J}^{\mathrm{ra}} =&f \int \mathrm{d} \mathbf{X}^{\prime} \mathbf{\Lambda}_{\mathbf{r}, \mathbf{X}^{\prime}}^{\mathrm{rt}} \hat{\mathbf{n}}^{\prime} \rho^{(2)}\left(\mathbf{X}, \mathbf{X}^{\prime}, t\right)
\end{aligned}
\end{equation}

Here $\rho^{(3)}(\mathbf{X}, \mathbf{X}', \mathbf{X}'', t)$ is the nonequilibrium triplet density. \\


We close these exact equation approximatively by using the DDFT relations on the pair and triplet level

\begin{equation}
\int \mathrm{d} \mathbf{r}^{\prime} \mathrm{d} \hat{\mathbf{n}}^{\prime} \rho^{(2)}\left(\mathbf{r}, \mathbf{r}^{\prime}, \hat{\mathbf{n}}, \hat{\mathbf{n}}^{\prime} ,t\right) \nabla_{\mathbf{r}^{\prime}} v\left(|\mathbf{r}-\mathbf{r}^{\prime}|\right)=\rho^{(1)}(\mathbf{r}, \hat{\mathbf{n}}, t) \nabla_{\mathbf{r}} \frac{\delta \mathcal{F}_{\mathrm{e} x c}}{\delta \rho^{(1)}(\mathbf{r}, \hat{\mathbf{n}}, t)}
\end{equation}

\begin{align*}
&\nabla_{\mathbf{r}^{\prime}} \rho^{(2)}\left(\mathbf{r}, \mathbf{r}^{\prime} \hat{\mathbf{n}}, \hat{\mathbf{n}}^{\prime} t\right)+\rho^{(2)}\left(\mathbf{r}, \mathbf{r}^{\prime} \hat{\mathbf{n}}, \hat{\mathbf{n}}^{\prime}, t\right) \nabla_{\mathbf{r}^{\prime}} v\left(|\mathbf{r}-\mathbf{r}^{\prime}|\right) \\
&+\int \mathrm{d} \mathbf{r}^{\prime} \mathrm{d} \hat{\mathbf{n}}^{\prime} \rho^{(3)}\left(\mathbf{r}, \mathbf{r}^{\prime}, \mathbf{r}^{\prime \prime}, \hat{\mathbf{n}}, \hat{\mathbf{n}}^{\prime}, \hat{\mathbf{n}}^{\prime \prime}, t\right) \nabla_{\mathbf{r}^{\prime}} u\left(\mathbf{r}^{\prime} \mathbf{r}^{\prime \prime}\right) \\
&=\rho^{(2)}\left(\mathbf{r}, \mathbf{r}^{\prime} \hat{\mathbf{n}}, \hat{\mathbf{n}}^{\prime}, t\right)\left(\nabla_{\mathbf{r}^{\prime}} \ln \left(\lambda^{3} \rho^{(1)}\left(\mathbf{r}^{\prime}, \hat{\mathbf{n}}^{\prime}, t\right)\right)+\nabla_{\mathbf{r}^{\prime}} \frac{\delta \mathcal{F}_{e x c}}{\delta \rho^{(1)}\left(\mathbf{r}^{\prime}, \hat{\mathbf{n}}^{\prime}, t\right)}\right)
\end{align*}


The remaining input is a concrete approximation for the equilibrium density functional
 where we adopt the mean-field approximation
\begin{equation}
\mathcal{F}_{e x c}=\frac{1}{2} \int \text{d}\mathbf r \text{d}\mathbf r'\;\text{d}\hat{\mathbf n} \text{d}\hat{\mathbf n}'  \rho^{(1)}(\mathbf{r}, \hat{\mathbf{n}}, t) \rho^{(1)}\left(\mathbf{r}^{\prime}, \hat{\mathbf{n}}^{\prime}, t\right) v(|\mathbf{r} - \mathbf{r}'|)
\end{equation}
and a high-temperature factorization approximation for the remaining nonequilibrium pair correlation 
\begin{equation}
\rho^{(2)}\left(\mathbf{r},{\mathbf{r}}^{\prime}, \hat{\mathbf{n}}, \hat{\mathbf{n}}^{\prime}, t\right)=\rho^{(1)}(\mathbf{r}, \hat{\mathbf{n}}, t) \rho^{(1)}\left(\mathbf{r}^{\prime}, \hat{\mathbf{n}}^{\prime}, t\right) \exp \left(-\beta v(|\mathbf{r} - \mathbf{r}'|)\right)
\end{equation}
Then the full set of equations is closed. They only involve the dynamical one-body density field and can be solved numerically.
\end{widetext}

\subsection{Applications of DDFT to microswimmers}

As a first application, we consider the motion of microswimmers which are moving in two spatial dimensions surrounded by a three-dimensional bulk fluid \cite{Menzel_JChemPhys_2016}. They are confined to an quartic external potential
\begin{equation}
V_{ext}(\mathbf r)=k|\mathbf r|^4.
\end{equation} 
with $k$ defining a confinement strength
and exhibit non-aligning interactions embodied in the steric pair potential
\begin{equation}
v(r)=\epsilon_0\exp\left(-\frac{r^4}{\sigma^4}\right). 
\label{eqGEM4}
\end{equation}
here, $\epsilon_0$ sets the strength of this potential and $\sigma$ an effective interaction range. 

The calculation protocol is to turn the activity $f$ off first and equilibrate the particles in the quartic potential. The parameters are chosen in such a way that the equilibrium system is in the fluid phase but exhibits density peaks due to the steric potential. Then the self-propulsion $f$ is switched on and the density evolution is followed by solving the DDFT equations numerically. For small self-propulsion strengths $|f|$,
 a stationary high-density ring is formed both for pushers ($f>0$) and pullers ($f<0$) which is extended relative to the typical extension of the equilibrated system. 
If the self-propulsion strength is getting larger a tangential instability occurs and the system forms spontaneously a hydrodynamic pump
as predicted earlier by simulations \cite{Ref_80_von_Menzel_JChemPhys_2016,Ref_81_von_Menzel_JChemPhys_2016}.
For even higher $|f|$ the pump gets unstable resulting in a continuous dynamic "swashing" of the density cloud.
The behaviour is similar for pushers and pullers but details are different, see Figure 2.


\begin{figure*}
\includegraphics[width=0.85\textwidth]{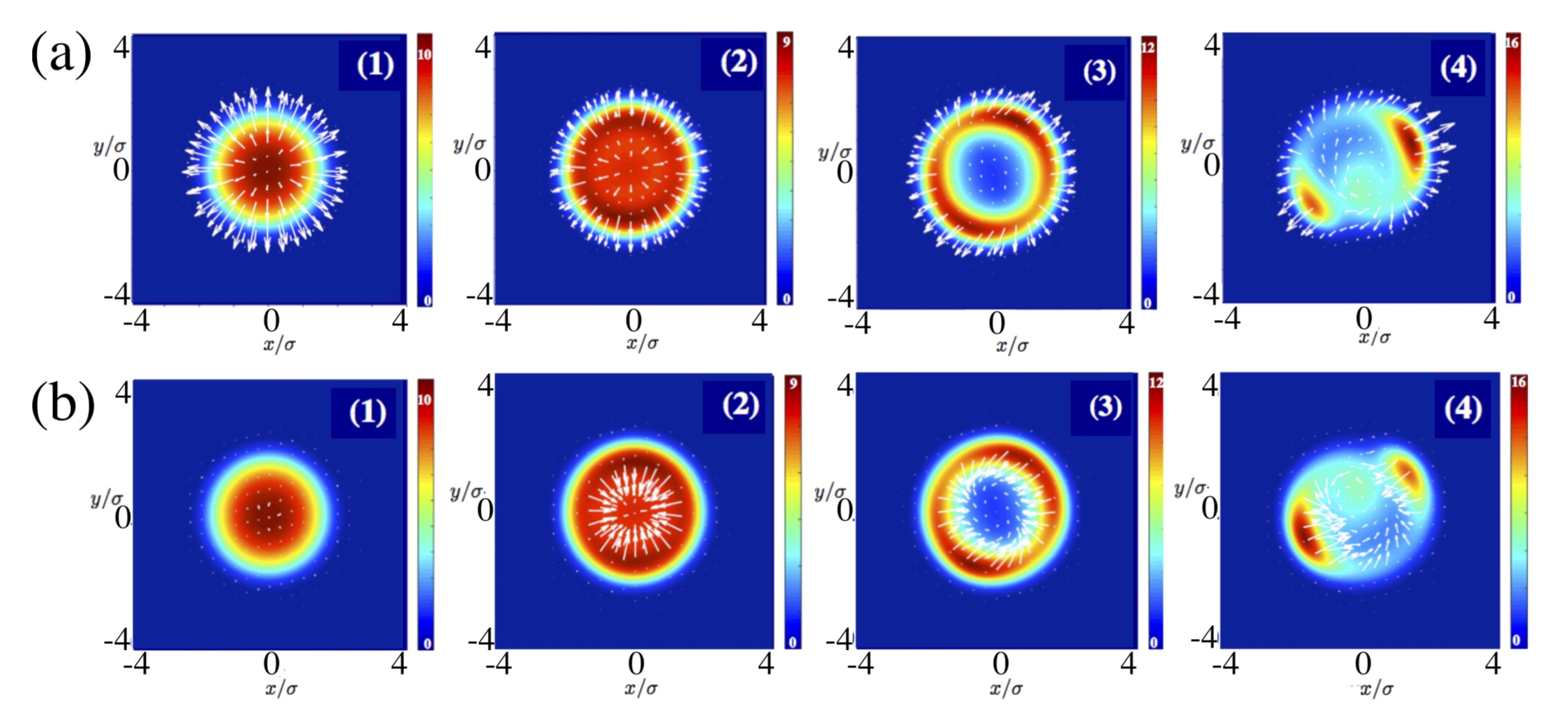}
\caption{ Time evolution of the density profiles (color maps) and orientation profiles (white arrows) (a) for pushers and (b) for pullers. The snapshots were obtained at times $t = 0.05, 0.1, 0.2, 0.8$. From Ref. \cite{Menzel_JChemPhys_2016}}
\label{Fig4MenzelJCPM}
\end{figure*}


The swimmer model can be generalized to force centers which are not collinear with the swimmer.
 This results in circe-swimming which was further analyzed within DDFT in \cite{Hoell_circle_swimmers}.
Moreover the DDFT approach was applied to global polar ordering in pure pusher or puller suspensions 
\cite{Hoell_polar_ordering}. As a result, at sufficient high concentrations polar ordering was found for pullers but not for pushers.
Finally the versatility of the DDFT is documented by its generalization to binary mixtures and to dynamics in a sheared fluid
which were considered and elaborated in Ref.\  \cite{Hoell_binary_DDFT}. \\

Let us finish with a remark: If one has dry active matter in mind from the very beginning, it is better to start with the approach described in Section V. The limit of vanishing hydrodynamic interactions is not a simple one if one does uses the swimmer model of this section, since hydrodynamics and self-propulsion are intrinsically coupled here. 

\section{Conclusions}

In conclusion, dynamical density functional theory which has been known as a successful theory for interacting Brownian colloidal particles 
can be applied to active matter as well. In particular, both "dry active matter" and "wet active matter" (microswimmers) 
can be treated on different levels of complexity.  

Future research will be directed along the following topics: 

i) Density functional theory provides an ideal framework to tackle {\it aligning interactions}. This strength needs to be exploited further to
establish a first-principle approach to the Vicsek-model of swarming \cite{vicsek1995novel}  and to the impact of alignment effects 
on motility-induced phase separation \cite{Nilsson2017,Sese_Sansa2018,Dijkstra2020}.

ii) So far we discussed swimmers in a viscous Newtonian fluid, but in many situations there is a {\it viscoelastic solvent}. Then memory effects of the 
solvent play a role which modify and affect the swimming process. One basic example for a viscoelastic medium is a polymer solution. 
For colloids swimming in a polymer solution, a strongly enhanced rotational diffusion was found in experiment \cite{Bechinger_exp} 
and simulation \cite{Gompper_sim}. 
It is challenging to treat these memory effects of the viscoelastic solvent within dynamical density functional theory.

iii) Density functional theory is ideal for the calculation of {\it interfaces and wetting effects} \cite{Evans_Ann}. 
So it should be applied to study interfaces between coexisting states for active particles.
 This can be both fluid-fluid and fluid-solid interfaces. 
 For an effective equilibrium model for dry active particles, 
this was done by Wittmann and Brader \cite{wetting} but an extension towards the full DDFT is still needed.

iv) As density functional theory describes freezing and crystallization 
in equilibrium, dynamical density functional theory should  be applied to {\it freezing} of active particles \cite{Speck_2012_PRL}.
A simplified approach based on the phase-field-crystal model has been proposed \cite{Menzel_2013_PRL}  but this needs extension towards
a theory which includes microscopic correlations. 

v) Particles with {\it time-dependent pair interactions} such as breathing particles whose interaction diameter changes periodically as a function of time.
There should be  no principle obstacle to formulate a DDFT for these nonequilibrium systems which play an important role for modelling 
dense biological tissue \cite{Berthier2017}.

vi) Bacteria subjected to simultaneous {\it growth and division} establish a  complex dynamical phenomenon when strongly interacting \cite{Idema}.
A DDFT approach seems to be particularly promising to described nematic order on the particle-scale for growing bacterial colonies \cite{Blow_Poon}.

\section{Exercises}

\textbf{Exercise 1}: 
Calculate the first two functional derivatives $\frac{\delta \mathcal{F}[\rho]}{\delta \rho(\mathbf{r})}$ and $\frac{\delta^2 \mathcal{F}[\rho]}{\delta \rho(\mathbf{r}) \, \delta \rho(\mathbf{r}\,')}$ for 
\begin{enumerate}
	\item $\mathcal{F}[\rho] = \frac12 \intdr_1 \intdr_2 \, w(|\mathbf{r}_1 - \mathbf{r}_2|) \rho(\mathbf{r}_1) \rho(\mathbf{r}_2)$,
	\item $\mathcal{F}[\rho] = \intdr_1 \Psi(\rho(\mathbf{r}_1))$
\end{enumerate}
in three spatial dimensions. \\
Here, $w(r)$ and $\Psi(\rho )$ are prescribed given functions.

\textbf{Exercise 2}:
Show that for the ideal gas in an external potential the dynamical density functional theory
reduces to the exact Smoluchowski equation.

\textbf{Exercise 3}: Show that the DDFT for dry active matter is equivalent to the underlying Smoluchowski equation
is the particles are non-interacting ($v(r)=0$) but exposed to a general external potential $V_{\text{ext}}({\mathbf r}, {\hat{ \mathbf{n}}},t)$.
By the way it has been erroneously claimed in the literature that DDFT is approximative in this case \cite{Enculescu_Stark}.  

\textbf{Exercise 4}: Give the DDFT equations for the dynamics of the density field 
$\rho(\mathbf{r},\phi,t)$ for dry active matter with the mean-field approximation of the density
 functional and aligning interactions. \\
 
\textbf{Acknowledgement}: I thank R. Wittkowski, C. Hoell, A. M. Menzel and R. Wittmann
 for many helpful suggestions. This work was supported by
the MSCA-ITN ActiveMatter (EU proposal 812780).\\[0,2cm]

\bibliography{bibliography}

\begin{thebibliography}{51}%
\makeatletter
\providecommand \@ifxundefined [1]{%
 \@ifx{#1\undefined}
}%
\providecommand \@ifnum [1]{%
 \ifnum #1\expandafter \@firstoftwo
 \else \expandafter \@secondoftwo
 \fi
}%
\providecommand \@ifx [1]{%
 \ifx #1\expandafter \@firstoftwo
 \else \expandafter \@secondoftwo
 \fi
}%
\providecommand \natexlab [1]{#1}%
\providecommand \enquote  [1]{``#1''}%
\providecommand \bibnamefont  [1]{#1}%
\providecommand \bibfnamefont [1]{#1}%
\providecommand \citenamefont [1]{#1}%
\providecommand \href@noop [0]{\@secondoftwo}%
\providecommand \href [0]{\begingroup \@sanitize@url \@href}%
\providecommand \@href[1]{\@@startlink{#1}\@@href}%
\providecommand \@@href[1]{\endgroup#1\@@endlink}%
\providecommand \@sanitize@url [0]{\catcode `\\12\catcode `\$12\catcode
  `\&12\catcode `\#12\catcode `\^12\catcode `\_12\catcode `\%12\relax}%
\providecommand \@@startlink[1]{}%
\providecommand \@@endlink[0]{}%
\providecommand \url  [0]{\begingroup\@sanitize@url \@url }%
\providecommand \@url [1]{\endgroup\@href {#1}{\urlprefix }}%
\providecommand \urlprefix  [0]{URL }%
\providecommand \Eprint [0]{\href }%
\providecommand \doibase [0]{http://dx.doi.org/}%
\providecommand \selectlanguage [0]{\@gobble}%
\providecommand \bibinfo  [0]{\@secondoftwo}%
\providecommand \bibfield  [0]{\@secondoftwo}%
\providecommand \translation [1]{[#1]}%
\providecommand \BibitemOpen [0]{}%
\providecommand \bibitemStop [0]{}%
\providecommand \bibitemNoStop [0]{.\EOS\space}%
\providecommand \EOS [0]{\spacefactor3000\relax}%
\providecommand \BibitemShut  [1]{\csname bibitem#1\endcsname}%
\let\auto@bib@innerbib\@empty
\bibitem [{\citenamefont {L{\"o}wen}(2017)}]{Springer_book}%
  \BibitemOpen
  \bibfield  {author} {\bibinfo {author} {\bibfnamefont {H.}~\bibnamefont
  {L{\"o}wen}},\ }in\ \href@noop {} {\emph {\bibinfo {booktitle} {Variational
  Methods in Molecular Modeling}}}\ (\bibinfo  {publisher} {Springer},\
  \bibinfo {year} {2017})\ pp.\ \bibinfo {pages} {255--284}\BibitemShut
  {NoStop}%
\bibitem [{\citenamefont {te~Vrugt}\ \emph {et~al.}(2020)\citenamefont
  {te~Vrugt}, \citenamefont {L{\"o}wen},\ and\ \citenamefont
  {Wittkowski}}]{review_DDFT}%
  \BibitemOpen
  \bibfield  {author} {\bibinfo {author} {\bibfnamefont {M.}~\bibnamefont
  {te~Vrugt}}, \bibinfo {author} {\bibfnamefont {H.}~\bibnamefont {L{\"o}wen}},
  \ and\ \bibinfo {author} {\bibfnamefont {R.}~\bibnamefont {Wittkowski}},\
  }\href@noop {} {\bibfield  {journal} {\bibinfo  {journal} {Adv. in Phys.}\
  }\textbf {\bibinfo {volume} {69}},\ \bibinfo {pages} {121} (\bibinfo {year}
  {2020})}\BibitemShut {NoStop}%
\bibitem [{\citenamefont {L{\"o}wen}(1994)}]{mein_review}%
  \BibitemOpen
  \bibfield  {author} {\bibinfo {author} {\bibfnamefont {H.}~\bibnamefont
  {L{\"o}wen}},\ }\href@noop {} {\bibfield  {journal} {\bibinfo  {journal}
  {Physics Reports}\ }\textbf {\bibinfo {volume} {237}},\ \bibinfo {pages}
  {249} (\bibinfo {year} {1994})}\BibitemShut {NoStop}%
\bibitem [{\citenamefont {Roth}(2010)}]{Roth_review_JPCM_2010}%
  \BibitemOpen
  \bibfield  {author} {\bibinfo {author} {\bibfnamefont {R.}~\bibnamefont
  {Roth}},\ }\href@noop {} {\bibfield  {journal} {\bibinfo  {journal} {J. Phys.
  Condensed Matter}\ }\textbf {\bibinfo {volume} {22}},\ \bibinfo {pages}
  {063102} (\bibinfo {year} {2010})}\BibitemShut {NoStop}%
\bibitem [{\citenamefont {Roth}\ \emph {et~al.}(2012)\citenamefont {Roth},
  \citenamefont {Mecke},\ and\ \citenamefont {Oettel}}]{Oettel_JCP}%
  \BibitemOpen
  \bibfield  {author} {\bibinfo {author} {\bibfnamefont {R.}~\bibnamefont
  {Roth}}, \bibinfo {author} {\bibfnamefont {K.}~\bibnamefont {Mecke}}, \ and\
  \bibinfo {author} {\bibfnamefont {M.}~\bibnamefont {Oettel}},\ }\href@noop {}
  {\bibfield  {journal} {\bibinfo  {journal} {J. Chem. Phys.}\ }\textbf
  {\bibinfo {volume} {136}},\ \bibinfo {pages} {081101} (\bibinfo {year}
  {2012})}\BibitemShut {NoStop}%
\bibitem [{\citenamefont {Doi}\ and\ \citenamefont
  {Edwards}(1988)}]{Doi_Edwards}%
  \BibitemOpen
  \bibfield  {author} {\bibinfo {author} {\bibfnamefont {M.}~\bibnamefont
  {Doi}}\ and\ \bibinfo {author} {\bibfnamefont {S.~F.}\ \bibnamefont
  {Edwards}},\ }\href@noop {} {\emph {\bibinfo {title} {{The Theory of Polymer
  Dynamics}}}},\ Vol.~\bibinfo {volume} {73}\ (\bibinfo  {publisher} {oxford
  university press},\ \bibinfo {year} {1988})\BibitemShut {NoStop}%
\bibitem [{\citenamefont {Archer}\ and\ \citenamefont
  {Evans}(2004)}]{Archer_Evans}%
  \BibitemOpen
  \bibfield  {author} {\bibinfo {author} {\bibfnamefont {A.~J.}\ \bibnamefont
  {Archer}}\ and\ \bibinfo {author} {\bibfnamefont {R.}~\bibnamefont {Evans}},\
  }\href@noop {} {\bibfield  {journal} {\bibinfo  {journal} {J. Chem. Phys.}\
  }\textbf {\bibinfo {volume} {121}},\ \bibinfo {pages} {4246} (\bibinfo {year}
  {2004})}\BibitemShut {NoStop}%
\bibitem [{\citenamefont {Espa{\~n}ol}\ and\ \citenamefont
  {L{\"o}wen}(2009)}]{JCP_2009}%
  \BibitemOpen
  \bibfield  {author} {\bibinfo {author} {\bibfnamefont {P.}~\bibnamefont
  {Espa{\~n}ol}}\ and\ \bibinfo {author} {\bibfnamefont {H.}~\bibnamefont
  {L{\"o}wen}},\ }\href@noop {} {\bibfield  {journal} {\bibinfo  {journal} {J.
  Chem. Phys.}\ }\textbf {\bibinfo {volume} {131}},\ \bibinfo {pages} {244101}
  (\bibinfo {year} {2009})}\BibitemShut {NoStop}%
\bibitem [{\citenamefont {Marconi}\ and\ \citenamefont
  {Tarazona}(1999)}]{Marconi_1}%
  \BibitemOpen
  \bibfield  {author} {\bibinfo {author} {\bibfnamefont {U.~M.~B.}\
  \bibnamefont {Marconi}}\ and\ \bibinfo {author} {\bibfnamefont
  {P.}~\bibnamefont {Tarazona}},\ }\href@noop {} {\bibfield  {journal}
  {\bibinfo  {journal} {J. Chem. Phys.}\ }\textbf {\bibinfo {volume} {110}},\
  \bibinfo {pages} {8032} (\bibinfo {year} {1999})}\BibitemShut {NoStop}%
\bibitem [{\citenamefont {Marconi}\ and\ \citenamefont
  {Tarazona}(2000)}]{Marconi_2}%
  \BibitemOpen
  \bibfield  {author} {\bibinfo {author} {\bibfnamefont {U.~M.~B.}\
  \bibnamefont {Marconi}}\ and\ \bibinfo {author} {\bibfnamefont
  {P.}~\bibnamefont {Tarazona}},\ }\href@noop {} {\bibfield  {journal}
  {\bibinfo  {journal} {J. Phys. Condens. Matter}\ }\textbf {\bibinfo {volume}
  {12}},\ \bibinfo {pages} {A413} (\bibinfo {year} {2000})}\BibitemShut
  {NoStop}%
\bibitem [{\citenamefont {Wittmann}\ \emph {et~al.}(2017)\citenamefont
  {Wittmann}, \citenamefont {Sitta}, \citenamefont {Smallenburg},\ and\
  \citenamefont {L{\"o}wen}}]{Wittmann}%
  \BibitemOpen
  \bibfield  {author} {\bibinfo {author} {\bibfnamefont {R.}~\bibnamefont
  {Wittmann}}, \bibinfo {author} {\bibfnamefont {C.~E.}\ \bibnamefont {Sitta}},
  \bibinfo {author} {\bibfnamefont {F.}~\bibnamefont {Smallenburg}}, \ and\
  \bibinfo {author} {\bibfnamefont {H.}~\bibnamefont {L{\"o}wen}},\ }\href@noop
  {} {\bibfield  {journal} {\bibinfo  {journal} {J. Chem. Phys.}\ }\textbf
  {\bibinfo {volume} {147}},\ \bibinfo {pages} {134908} (\bibinfo {year}
  {2017})}\BibitemShut {NoStop}%
\bibitem [{\citenamefont {Rex}\ \emph {et~al.}(2007)\citenamefont {Rex},
  \citenamefont {Wensink},\ and\ \citenamefont {L\"owen}}]{Rex_Wensink_2007}%
  \BibitemOpen
  \bibfield  {author} {\bibinfo {author} {\bibfnamefont {M.}~\bibnamefont
  {Rex}}, \bibinfo {author} {\bibfnamefont {H.~H.}\ \bibnamefont {Wensink}}, \
  and\ \bibinfo {author} {\bibfnamefont {H.}~\bibnamefont {L\"owen}},\
  }\href@noop {} {\bibfield  {journal} {\bibinfo  {journal} {Phys. Rev. E}\
  }\textbf {\bibinfo {volume} {76}},\ \bibinfo {pages} {021403} (\bibinfo
  {year} {2007})}\BibitemShut {NoStop}%
\bibitem [{\citenamefont {ten Hagen}\ \emph {et~al.}(2015)\citenamefont {ten
  Hagen}, \citenamefont {Wittkowski}, \citenamefont {Takagi}, \citenamefont
  {K{\"u}mmel}, \citenamefont {Bechinger},\ and\ \citenamefont
  {L{\"o}wen}}]{JPCM_2015}%
  \BibitemOpen
  \bibfield  {author} {\bibinfo {author} {\bibfnamefont {B.}~\bibnamefont {ten
  Hagen}}, \bibinfo {author} {\bibfnamefont {R.}~\bibnamefont {Wittkowski}},
  \bibinfo {author} {\bibfnamefont {D.}~\bibnamefont {Takagi}}, \bibinfo
  {author} {\bibfnamefont {F.}~\bibnamefont {K{\"u}mmel}}, \bibinfo {author}
  {\bibfnamefont {C.}~\bibnamefont {Bechinger}}, \ and\ \bibinfo {author}
  {\bibfnamefont {H.}~\bibnamefont {L{\"o}wen}},\ }\href@noop {} {\bibfield
  {journal} {\bibinfo  {journal} {J. Phys. Condens. Matter}\ }\textbf {\bibinfo
  {volume} {27}},\ \bibinfo {pages} {194110} (\bibinfo {year}
  {2015})}\BibitemShut {NoStop}%
\bibitem [{\citenamefont {Howse}\ \emph {et~al.}(2007)\citenamefont {Howse},
  \citenamefont {Jones}, \citenamefont {Ryan}, \citenamefont {Gough},
  \citenamefont {Vafabakhsh},\ and\ \citenamefont {Golestanian}}]{Golestanian}%
  \BibitemOpen
  \bibfield  {author} {\bibinfo {author} {\bibfnamefont {J.~R.}\ \bibnamefont
  {Howse}}, \bibinfo {author} {\bibfnamefont {R.~A.}\ \bibnamefont {Jones}},
  \bibinfo {author} {\bibfnamefont {A.~J.}\ \bibnamefont {Ryan}}, \bibinfo
  {author} {\bibfnamefont {T.}~\bibnamefont {Gough}}, \bibinfo {author}
  {\bibfnamefont {R.}~\bibnamefont {Vafabakhsh}}, \ and\ \bibinfo {author}
  {\bibfnamefont {R.}~\bibnamefont {Golestanian}},\ }\href@noop {} {\bibfield
  {journal} {\bibinfo  {journal} {Phys. Rev. Lett.}\ }\textbf {\bibinfo
  {volume} {99}},\ \bibinfo {pages} {048102} (\bibinfo {year}
  {2007})}\BibitemShut {NoStop}%
\bibitem [{\citenamefont {van Teeffelen}\ and\ \citenamefont
  {L\"owen}(2008)}]{Teeffelen_PRE_2008}%
  \BibitemOpen
  \bibfield  {author} {\bibinfo {author} {\bibfnamefont {S.}~\bibnamefont {van
  Teeffelen}}\ and\ \bibinfo {author} {\bibfnamefont {H.}~\bibnamefont
  {L\"owen}},\ }\href@noop {} {\bibfield  {journal} {\bibinfo  {journal} {Phys.
  Rev. E}\ }\textbf {\bibinfo {volume} {78}},\ \bibinfo {pages} {020101}
  (\bibinfo {year} {2008})}\BibitemShut {NoStop}%
\bibitem [{\citenamefont {ten Hagen}\ \emph {et~al.}(2011)\citenamefont {ten
  Hagen}, \citenamefont {van Teeffelen},\ and\ \citenamefont
  {L{\"o}wen}}]{Hagen_JPCM}%
  \BibitemOpen
  \bibfield  {author} {\bibinfo {author} {\bibfnamefont {B.}~\bibnamefont {ten
  Hagen}}, \bibinfo {author} {\bibfnamefont {S.}~\bibnamefont {van Teeffelen}},
  \ and\ \bibinfo {author} {\bibfnamefont {H.}~\bibnamefont {L{\"o}wen}},\
  }\href@noop {} {\bibfield  {journal} {\bibinfo  {journal} {J. Phys. Condens.
  Matter}\ }\textbf {\bibinfo {volume} {23}},\ \bibinfo {pages} {194119}
  (\bibinfo {year} {2011})}\BibitemShut {NoStop}%
\bibitem [{\citenamefont {Bechinger}\ \emph {et~al.}(2016)\citenamefont
  {Bechinger}, \citenamefont {Di~Leonardo}, \citenamefont {L{\"o}wen},
  \citenamefont {Reichhardt}, \citenamefont {Volpe},\ and\ \citenamefont
  {Volpe}}]{our_RMP}%
  \BibitemOpen
  \bibfield  {author} {\bibinfo {author} {\bibfnamefont {C.}~\bibnamefont
  {Bechinger}}, \bibinfo {author} {\bibfnamefont {R.}~\bibnamefont
  {Di~Leonardo}}, \bibinfo {author} {\bibfnamefont {H.}~\bibnamefont
  {L{\"o}wen}}, \bibinfo {author} {\bibfnamefont {C.}~\bibnamefont
  {Reichhardt}}, \bibinfo {author} {\bibfnamefont {G.}~\bibnamefont {Volpe}}, \
  and\ \bibinfo {author} {\bibfnamefont {G.}~\bibnamefont {Volpe}},\
  }\href@noop {} {\bibfield  {journal} {\bibinfo  {journal} {Rev. Mod. Phys.}\
  }\textbf {\bibinfo {volume} {88}},\ \bibinfo {pages} {045006} (\bibinfo
  {year} {2016})}\BibitemShut {NoStop}%
\bibitem [{\citenamefont {Kurzthaler}\ \emph {et~al.}(2016)\citenamefont
  {Kurzthaler}, \citenamefont {Leitmann},\ and\ \citenamefont
  {Franosch}}]{Kurzthaler_2016}%
  \BibitemOpen
  \bibfield  {author} {\bibinfo {author} {\bibfnamefont {C.}~\bibnamefont
  {Kurzthaler}}, \bibinfo {author} {\bibfnamefont {S.}~\bibnamefont
  {Leitmann}}, \ and\ \bibinfo {author} {\bibfnamefont {T.}~\bibnamefont
  {Franosch}},\ }\href@noop {} {\bibfield  {journal} {\bibinfo  {journal} {Sci.
  Rep.}\ }\textbf {\bibinfo {volume} {6}},\ \bibinfo {pages} {36702} (\bibinfo
  {year} {2016})}\BibitemShut {NoStop}%
\bibitem [{\citenamefont {Wensink}\ and\ \citenamefont
  {L{\"o}wen}(2008)}]{Wensink_2008}%
  \BibitemOpen
  \bibfield  {author} {\bibinfo {author} {\bibfnamefont {H.}~\bibnamefont
  {Wensink}}\ and\ \bibinfo {author} {\bibfnamefont {H.}~\bibnamefont
  {L{\"o}wen}},\ }\href@noop {} {\bibfield  {journal} {\bibinfo  {journal}
  {Phys. Rev. E}\ }\textbf {\bibinfo {volume} {78}},\ \bibinfo {pages} {031409}
  (\bibinfo {year} {2008})}\BibitemShut {NoStop}%
\bibitem [{\citenamefont {Enculescu}\ and\ \citenamefont
  {Stark}(2011)}]{Enculescu_Stark}%
  \BibitemOpen
  \bibfield  {author} {\bibinfo {author} {\bibfnamefont {M.}~\bibnamefont
  {Enculescu}}\ and\ \bibinfo {author} {\bibfnamefont {H.}~\bibnamefont
  {Stark}},\ }\href@noop {} {\bibfield  {journal} {\bibinfo  {journal} {Phys.
  Rev. Lett.}\ }\textbf {\bibinfo {volume} {107}},\ \bibinfo {pages} {058301}
  (\bibinfo {year} {2011})}\BibitemShut {NoStop}%
\bibitem [{\citenamefont {ten Hagen}\ \emph {et~al.}(2014)\citenamefont {ten
  Hagen}, \citenamefont {K{\"u}mmel}, \citenamefont {Wittkowski}, \citenamefont
  {Takagi}, \citenamefont {L{\"o}wen},\ and\ \citenamefont
  {Bechinger}}]{Kuemmel}%
  \BibitemOpen
  \bibfield  {author} {\bibinfo {author} {\bibfnamefont {B.}~\bibnamefont {ten
  Hagen}}, \bibinfo {author} {\bibfnamefont {F.}~\bibnamefont {K{\"u}mmel}},
  \bibinfo {author} {\bibfnamefont {R.}~\bibnamefont {Wittkowski}}, \bibinfo
  {author} {\bibfnamefont {D.}~\bibnamefont {Takagi}}, \bibinfo {author}
  {\bibfnamefont {H.}~\bibnamefont {L{\"o}wen}}, \ and\ \bibinfo {author}
  {\bibfnamefont {C.}~\bibnamefont {Bechinger}},\ }\href@noop {} {\bibfield
  {journal} {\bibinfo  {journal} {{Nature Commun.}}\ }\textbf {\bibinfo
  {volume} {5}},\ \bibinfo {pages} {4829} (\bibinfo {year} {2014})}\BibitemShut
  {NoStop}%
\bibitem [{\citenamefont {Hermann}\ and\ \citenamefont
  {Schmidt}(2018)}]{Herrmann_Schmidt}%
  \BibitemOpen
  \bibfield  {author} {\bibinfo {author} {\bibfnamefont {S.}~\bibnamefont
  {Hermann}}\ and\ \bibinfo {author} {\bibfnamefont {M.}~\bibnamefont
  {Schmidt}},\ }\href@noop {} {\bibfield  {journal} {\bibinfo  {journal} {Soft
  Matter}\ }\textbf {\bibinfo {volume} {14}},\ \bibinfo {pages} {1614}
  (\bibinfo {year} {2018})}\BibitemShut {NoStop}%
\bibitem [{\citenamefont {Vachier}\ and\ \citenamefont {Mazza}(2019)}]{Mazza}%
  \BibitemOpen
  \bibfield  {author} {\bibinfo {author} {\bibfnamefont {J.}~\bibnamefont
  {Vachier}}\ and\ \bibinfo {author} {\bibfnamefont {M.~G.}\ \bibnamefont
  {Mazza}},\ }\href@noop {} {\bibfield  {journal} {\bibinfo  {journal} {Eur.
  Phys. J. E}\ }\textbf {\bibinfo {volume} {42}},\ \bibinfo {pages} {11}
  (\bibinfo {year} {2019})}\BibitemShut {NoStop}%
\bibitem [{\citenamefont {Onsager}(1949)}]{Onsager}%
  \BibitemOpen
  \bibfield  {author} {\bibinfo {author} {\bibfnamefont {L.}~\bibnamefont
  {Onsager}},\ }\href@noop {} {\bibfield  {journal} {\bibinfo  {journal} {Proc.
  New York Acad. Sci.}\ }\textbf {\bibinfo {volume} {51}},\ \bibinfo {pages}
  {627} (\bibinfo {year} {1949})}\BibitemShut {NoStop}%
\bibitem [{\citenamefont {Wittkowski}\ and\ \citenamefont
  {L{\"o}wen}(2011)}]{Witt_Mol_Phys}%
  \BibitemOpen
  \bibfield  {author} {\bibinfo {author} {\bibfnamefont {R.}~\bibnamefont
  {Wittkowski}}\ and\ \bibinfo {author} {\bibfnamefont {H.}~\bibnamefont
  {L{\"o}wen}},\ }\href@noop {} {\bibfield  {journal} {\bibinfo  {journal}
  {Mol. Phys}\ }\textbf {\bibinfo {volume} {109}},\ \bibinfo {pages} {2935}
  (\bibinfo {year} {2011})}\BibitemShut {NoStop}%
\bibitem [{\citenamefont {Krinninger}\ \emph {et~al.}(2016)\citenamefont
  {Krinninger}, \citenamefont {Schmidt},\ and\ \citenamefont
  {Brader}}]{Krinninger_2016}%
  \BibitemOpen
  \bibfield  {author} {\bibinfo {author} {\bibfnamefont {P.}~\bibnamefont
  {Krinninger}}, \bibinfo {author} {\bibfnamefont {M.}~\bibnamefont {Schmidt}},
  \ and\ \bibinfo {author} {\bibfnamefont {J.~M.}\ \bibnamefont {Brader}},\
  }\href@noop {} {\bibfield  {journal} {\bibinfo  {journal} {Phys. Rev. Lett.}\
  }\textbf {\bibinfo {volume} {117}},\ \bibinfo {pages} {208003} (\bibinfo
  {year} {2016})}\BibitemShut {NoStop}%
\bibitem [{\citenamefont {Krinninger}\ and\ \citenamefont
  {Schmidt}(2019)}]{Krinninger_2019}%
  \BibitemOpen
  \bibfield  {author} {\bibinfo {author} {\bibfnamefont {P.}~\bibnamefont
  {Krinninger}}\ and\ \bibinfo {author} {\bibfnamefont {M.}~\bibnamefont
  {Schmidt}},\ }\href@noop {} {\bibfield  {journal} {\bibinfo  {journal} {J.
  Chem. Phys.}\ }\textbf {\bibinfo {volume} {150}},\ \bibinfo {pages} {074112}
  (\bibinfo {year} {2019})}\BibitemShut {NoStop}%
\bibitem [{\citenamefont {Hermann}\ \emph {et~al.}(2019)\citenamefont
  {Hermann}, \citenamefont {Krinninger}, \citenamefont {de~las Heras},\ and\
  \citenamefont {Schmidt}}]{Hermann_2019}%
  \BibitemOpen
  \bibfield  {author} {\bibinfo {author} {\bibfnamefont {S.}~\bibnamefont
  {Hermann}}, \bibinfo {author} {\bibfnamefont {P.}~\bibnamefont {Krinninger}},
  \bibinfo {author} {\bibfnamefont {D.}~\bibnamefont {de~las Heras}}, \ and\
  \bibinfo {author} {\bibfnamefont {M.}~\bibnamefont {Schmidt}},\ }\href@noop
  {} {\bibfield  {journal} {\bibinfo  {journal} {Phys. Rev. E}\ }\textbf
  {\bibinfo {volume} {100}},\ \bibinfo {pages} {052604} (\bibinfo {year}
  {2019})}\BibitemShut {NoStop}%
\bibitem [{\citenamefont {Menzel}\ \emph {et~al.}(2016)\citenamefont {Menzel},
  \citenamefont {Saha}, \citenamefont {Hoell},\ and\ \citenamefont
  {L{\"o}wen}}]{Menzel_JChemPhys_2016}%
  \BibitemOpen
  \bibfield  {author} {\bibinfo {author} {\bibfnamefont {A.~M.}\ \bibnamefont
  {Menzel}}, \bibinfo {author} {\bibfnamefont {A.}~\bibnamefont {Saha}},
  \bibinfo {author} {\bibfnamefont {C.}~\bibnamefont {Hoell}}, \ and\ \bibinfo
  {author} {\bibfnamefont {H.}~\bibnamefont {L{\"o}wen}},\ }\href@noop {}
  {\bibfield  {journal} {\bibinfo  {journal} {J. Chem. Phys.}\ }\textbf
  {\bibinfo {volume} {144}},\ \bibinfo {pages} {024115} (\bibinfo {year}
  {2016})}\BibitemShut {NoStop}%
\bibitem [{\citenamefont {Purcell}(1977)}]{purcell1977life}%
  \BibitemOpen
  \bibfield  {author} {\bibinfo {author} {\bibfnamefont {E.~M.}\ \bibnamefont
  {Purcell}},\ }\href@noop {} {\bibfield  {journal} {\bibinfo  {journal} {Am.
  J. Phys.}\ }\textbf {\bibinfo {volume} {45}},\ \bibinfo {pages} {3} (\bibinfo
  {year} {1977})}\BibitemShut {NoStop}%
\bibitem [{\citenamefont {Fily}\ \emph {et~al.}(2012)\citenamefont {Fily},
  \citenamefont {Baskaran},\ and\ \citenamefont
  {Marchetti}}]{fily2012cooperative}%
  \BibitemOpen
  \bibfield  {author} {\bibinfo {author} {\bibfnamefont {Y.}~\bibnamefont
  {Fily}}, \bibinfo {author} {\bibfnamefont {A.}~\bibnamefont {Baskaran}}, \
  and\ \bibinfo {author} {\bibfnamefont {M.~C.}\ \bibnamefont {Marchetti}},\
  }\href@noop {} {\bibfield  {journal} {\bibinfo  {journal} {Soft Matter}\
  }\textbf {\bibinfo {volume} {8}},\ \bibinfo {pages} {3002} (\bibinfo {year}
  {2012})}\BibitemShut {NoStop}%
\bibitem [{\citenamefont {Baskaran}\ and\ \citenamefont
  {Marchetti}(2009)}]{baskaran2009statistical}%
  \BibitemOpen
  \bibfield  {author} {\bibinfo {author} {\bibfnamefont {A.}~\bibnamefont
  {Baskaran}}\ and\ \bibinfo {author} {\bibfnamefont {M.~C.}\ \bibnamefont
  {Marchetti}},\ }\href@noop {} {\bibfield  {journal} {\bibinfo  {journal}
  {Proc. Natl. Acad. Sci. USA}\ }\textbf {\bibinfo {volume} {106}},\ \bibinfo
  {pages} {15567} (\bibinfo {year} {2009})}\BibitemShut {NoStop}%
\bibitem [{\citenamefont {Happel}\ and\ \citenamefont
  {Brenner}(1981)}]{happel2012low}%
  \BibitemOpen
  \bibfield  {author} {\bibinfo {author} {\bibfnamefont {J.}~\bibnamefont
  {Happel}}\ and\ \bibinfo {author} {\bibfnamefont {H.}~\bibnamefont
  {Brenner}},\ }\href@noop {} {\emph {\bibinfo {title} {Low Reynolds number
  hydrodynamics}}}\ (\bibinfo  {publisher} {Springer Netherlands},\ \bibinfo
  {year} {1981})\BibitemShut {NoStop}%
\bibitem [{\citenamefont {Nash}\ \emph {et~al.}(2010)\citenamefont {Nash},
  \citenamefont {Adhikari}, \citenamefont {Tailleur},\ and\ \citenamefont
  {Cates}}]{Ref_80_von_Menzel_JChemPhys_2016}%
  \BibitemOpen
  \bibfield  {author} {\bibinfo {author} {\bibfnamefont {R.}~\bibnamefont
  {Nash}}, \bibinfo {author} {\bibfnamefont {R.}~\bibnamefont {Adhikari}},
  \bibinfo {author} {\bibfnamefont {J.}~\bibnamefont {Tailleur}}, \ and\
  \bibinfo {author} {\bibfnamefont {M.}~\bibnamefont {Cates}},\ }\href@noop {}
  {\bibfield  {journal} {\bibinfo  {journal} {Phys. Rev. Lett.}\ }\textbf
  {\bibinfo {volume} {104}},\ \bibinfo {pages} {258101} (\bibinfo {year}
  {2010})}\BibitemShut {NoStop}%
\bibitem [{\citenamefont {Hennes}\ \emph {et~al.}(2014)\citenamefont {Hennes},
  \citenamefont {Wolff},\ and\ \citenamefont
  {Stark}}]{Ref_81_von_Menzel_JChemPhys_2016}%
  \BibitemOpen
  \bibfield  {author} {\bibinfo {author} {\bibfnamefont {M.}~\bibnamefont
  {Hennes}}, \bibinfo {author} {\bibfnamefont {K.}~\bibnamefont {Wolff}}, \
  and\ \bibinfo {author} {\bibfnamefont {H.}~\bibnamefont {Stark}},\
  }\href@noop {} {\bibfield  {journal} {\bibinfo  {journal} {Phys. Rev. Lett.}\
  }\textbf {\bibinfo {volume} {112}},\ \bibinfo {pages} {238104} (\bibinfo
  {year} {2014})}\BibitemShut {NoStop}%
\bibitem [{\citenamefont {Hoell}\ \emph {et~al.}(2017)\citenamefont {Hoell},
  \citenamefont {L{\"o}wen},\ and\ \citenamefont
  {Menzel}}]{Hoell_circle_swimmers}%
  \BibitemOpen
  \bibfield  {author} {\bibinfo {author} {\bibfnamefont {C.}~\bibnamefont
  {Hoell}}, \bibinfo {author} {\bibfnamefont {H.}~\bibnamefont {L{\"o}wen}}, \
  and\ \bibinfo {author} {\bibfnamefont {A.~M.}\ \bibnamefont {Menzel}},\
  }\href@noop {} {\bibfield  {journal} {\bibinfo  {journal} {New J. Phys.}\
  }\textbf {\bibinfo {volume} {19}},\ \bibinfo {pages} {125004} (\bibinfo
  {year} {2017})}\BibitemShut {NoStop}%
\bibitem [{\citenamefont {Hoell}\ \emph {et~al.}(2018)\citenamefont {Hoell},
  \citenamefont {L{\"o}wen},\ and\ \citenamefont
  {Menzel}}]{Hoell_polar_ordering}%
  \BibitemOpen
  \bibfield  {author} {\bibinfo {author} {\bibfnamefont {C.}~\bibnamefont
  {Hoell}}, \bibinfo {author} {\bibfnamefont {H.}~\bibnamefont {L{\"o}wen}}, \
  and\ \bibinfo {author} {\bibfnamefont {A.~M.}\ \bibnamefont {Menzel}},\
  }\href@noop {} {\bibfield  {journal} {\bibinfo  {journal} {J. Chem. Phys.}\
  }\textbf {\bibinfo {volume} {149}},\ \bibinfo {pages} {144902} (\bibinfo
  {year} {2018})}\BibitemShut {NoStop}%
\bibitem [{\citenamefont {Hoell}\ \emph {et~al.}(2019)\citenamefont {Hoell},
  \citenamefont {L{\"o}wen},\ and\ \citenamefont {Menzel}}]{Hoell_binary_DDFT}%
  \BibitemOpen
  \bibfield  {author} {\bibinfo {author} {\bibfnamefont {C.}~\bibnamefont
  {Hoell}}, \bibinfo {author} {\bibfnamefont {H.}~\bibnamefont {L{\"o}wen}}, \
  and\ \bibinfo {author} {\bibfnamefont {A.~M.}\ \bibnamefont {Menzel}},\
  }\href@noop {} {\bibfield  {journal} {\bibinfo  {journal} {J. Chem. Phys.}\
  }\textbf {\bibinfo {volume} {151}},\ \bibinfo {pages} {064902} (\bibinfo
  {year} {2019})}\BibitemShut {NoStop}%
\bibitem [{\citenamefont {Vicsek}\ \emph {et~al.}(1995)\citenamefont {Vicsek},
  \citenamefont {Czir{\'o}k}, \citenamefont {Ben-Jacob}, \citenamefont
  {Cohen},\ and\ \citenamefont {Shochet}}]{vicsek1995novel}%
  \BibitemOpen
  \bibfield  {author} {\bibinfo {author} {\bibfnamefont {T.}~\bibnamefont
  {Vicsek}}, \bibinfo {author} {\bibfnamefont {A.}~\bibnamefont {Czir{\'o}k}},
  \bibinfo {author} {\bibfnamefont {E.}~\bibnamefont {Ben-Jacob}}, \bibinfo
  {author} {\bibfnamefont {I.}~\bibnamefont {Cohen}}, \ and\ \bibinfo {author}
  {\bibfnamefont {O.}~\bibnamefont {Shochet}},\ }\href@noop {} {\bibfield
  {journal} {\bibinfo  {journal} {Phys. Rev. Lett.}\ }\textbf {\bibinfo
  {volume} {75}},\ \bibinfo {pages} {1226} (\bibinfo {year}
  {1995})}\BibitemShut {NoStop}%
\bibitem [{\citenamefont {Nilsson}\ and\ \citenamefont
  {Volpe}(2017)}]{Nilsson2017}%
  \BibitemOpen
  \bibfield  {author} {\bibinfo {author} {\bibfnamefont {S.}~\bibnamefont
  {Nilsson}}\ and\ \bibinfo {author} {\bibfnamefont {G.}~\bibnamefont
  {Volpe}},\ }\href@noop {} {\bibfield  {journal} {\bibinfo  {journal} {New J.
  Phys.}\ }\textbf {\bibinfo {volume} {19}},\ \bibinfo {pages} {115008}
  (\bibinfo {year} {2017})}\BibitemShut {NoStop}%
\bibitem [{\citenamefont {Sese-Sansa}\ \emph {et~al.}(2018)\citenamefont
  {Sese-Sansa}, \citenamefont {Pagonabarraga},\ and\ \citenamefont
  {Levis}}]{Sese_Sansa2018}%
  \BibitemOpen
  \bibfield  {author} {\bibinfo {author} {\bibfnamefont {E.}~\bibnamefont
  {Sese-Sansa}}, \bibinfo {author} {\bibfnamefont {I.}~\bibnamefont
  {Pagonabarraga}}, \ and\ \bibinfo {author} {\bibfnamefont {D.}~\bibnamefont
  {Levis}},\ }\href@noop {} {\bibfield  {journal} {\bibinfo  {journal}
  {Europhys. Lett.}\ }\textbf {\bibinfo {volume} {124}},\ \bibinfo {pages}
  {30004} (\bibinfo {year} {2018})}\BibitemShut {NoStop}%
\bibitem [{\citenamefont {van~der Linden}\ \emph {et~al.}(2019)\citenamefont
  {van~der Linden}, \citenamefont {Alexander}, \citenamefont {Aarts},\ and\
  \citenamefont {Dauchot}}]{Dijkstra2020}%
  \BibitemOpen
  \bibfield  {author} {\bibinfo {author} {\bibfnamefont {M.~N.}\ \bibnamefont
  {van~der Linden}}, \bibinfo {author} {\bibfnamefont {L.~C.}\ \bibnamefont
  {Alexander}}, \bibinfo {author} {\bibfnamefont {D.~G.}\ \bibnamefont
  {Aarts}}, \ and\ \bibinfo {author} {\bibfnamefont {O.}~\bibnamefont
  {Dauchot}},\ }\href@noop {} {\bibfield  {journal} {\bibinfo  {journal} {Phys.
  Rev. Lett.}\ }\textbf {\bibinfo {volume} {123}},\ \bibinfo {pages} {098001}
  (\bibinfo {year} {2019})}\BibitemShut {NoStop}%
\bibitem [{\citenamefont {Gomez-Solano}\ \emph {et~al.}(2016)\citenamefont
  {Gomez-Solano}, \citenamefont {Blokhuis},\ and\ \citenamefont
  {Bechinger}}]{Bechinger_exp}%
  \BibitemOpen
  \bibfield  {author} {\bibinfo {author} {\bibfnamefont {J.~R.}\ \bibnamefont
  {Gomez-Solano}}, \bibinfo {author} {\bibfnamefont {A.}~\bibnamefont
  {Blokhuis}}, \ and\ \bibinfo {author} {\bibfnamefont {C.}~\bibnamefont
  {Bechinger}},\ }\href@noop {} {\bibfield  {journal} {\bibinfo  {journal}
  {Phys. Rev. Lett.}\ }\textbf {\bibinfo {volume} {116}},\ \bibinfo {pages}
  {138301} (\bibinfo {year} {2016})}\BibitemShut {NoStop}%
\bibitem [{\citenamefont {Qi}\ \emph {et~al.}(2020)\citenamefont {Qi},
  \citenamefont {Westphal}, \citenamefont {Gompper},\ and\ \citenamefont
  {Winkler}}]{Gompper_sim}%
  \BibitemOpen
  \bibfield  {author} {\bibinfo {author} {\bibfnamefont {K.}~\bibnamefont
  {Qi}}, \bibinfo {author} {\bibfnamefont {E.}~\bibnamefont {Westphal}},
  \bibinfo {author} {\bibfnamefont {G.}~\bibnamefont {Gompper}}, \ and\
  \bibinfo {author} {\bibfnamefont {R.~G.}\ \bibnamefont {Winkler}},\
  }\href@noop {} {\bibfield  {journal} {\bibinfo  {journal} {Phys. Rev. Lett.}\
  }\textbf {\bibinfo {volume} {124}},\ \bibinfo {pages} {068001} (\bibinfo
  {year} {2020})}\BibitemShut {NoStop}%
\bibitem [{\citenamefont {Evans}(1979)}]{Evans_Ann}%
  \BibitemOpen
  \bibfield  {author} {\bibinfo {author} {\bibfnamefont {R.}~\bibnamefont
  {Evans}},\ }\href@noop {} {\bibfield  {journal} {\bibinfo  {journal} {Adv.
  Phys.}\ }\textbf {\bibinfo {volume} {28}},\ \bibinfo {pages} {143} (\bibinfo
  {year} {1979})}\BibitemShut {NoStop}%
\bibitem [{\citenamefont {Wittmann}\ and\ \citenamefont
  {Brader}(2016)}]{wetting}%
  \BibitemOpen
  \bibfield  {author} {\bibinfo {author} {\bibfnamefont {R.}~\bibnamefont
  {Wittmann}}\ and\ \bibinfo {author} {\bibfnamefont {J.~M.}\ \bibnamefont
  {Brader}},\ }\href@noop {} {\bibfield  {journal} {\bibinfo  {journal}
  {Europhys. Lett.}\ }\textbf {\bibinfo {volume} {114}},\ \bibinfo {pages}
  {68004} (\bibinfo {year} {2016})}\BibitemShut {NoStop}%
\bibitem [{\citenamefont {Bialk{\'e}}\ \emph {et~al.}(2012)\citenamefont
  {Bialk{\'e}}, \citenamefont {Speck},\ and\ \citenamefont
  {L{\"o}wen}}]{Speck_2012_PRL}%
  \BibitemOpen
  \bibfield  {author} {\bibinfo {author} {\bibfnamefont {J.}~\bibnamefont
  {Bialk{\'e}}}, \bibinfo {author} {\bibfnamefont {T.}~\bibnamefont {Speck}}, \
  and\ \bibinfo {author} {\bibfnamefont {H.}~\bibnamefont {L{\"o}wen}},\
  }\href@noop {} {\bibfield  {journal} {\bibinfo  {journal} {Phys. Rev. Lett.}\
  }\textbf {\bibinfo {volume} {108}},\ \bibinfo {pages} {168301} (\bibinfo
  {year} {2012})}\BibitemShut {NoStop}%
\bibitem [{\citenamefont {Menzel}\ and\ \citenamefont
  {L{\"o}wen}(2013)}]{Menzel_2013_PRL}%
  \BibitemOpen
  \bibfield  {author} {\bibinfo {author} {\bibfnamefont {A.~M.}\ \bibnamefont
  {Menzel}}\ and\ \bibinfo {author} {\bibfnamefont {H.}~\bibnamefont
  {L{\"o}wen}},\ }\href@noop {} {\bibfield  {journal} {\bibinfo  {journal}
  {Phys. Rev. Lett.}\ }\textbf {\bibinfo {volume} {110}},\ \bibinfo {pages}
  {055702} (\bibinfo {year} {2013})}\BibitemShut {NoStop}%
\bibitem [{\citenamefont {Tjhung}\ and\ \citenamefont
  {Berthier}(2017)}]{Berthier2017}%
  \BibitemOpen
  \bibfield  {author} {\bibinfo {author} {\bibfnamefont {E.}~\bibnamefont
  {Tjhung}}\ and\ \bibinfo {author} {\bibfnamefont {L.}~\bibnamefont
  {Berthier}},\ }\href@noop {} {\bibfield  {journal} {\bibinfo  {journal}
  {Phys. Rev. E}\ }\textbf {\bibinfo {volume} {96}},\ \bibinfo {pages} {050601}
  (\bibinfo {year} {2017})}\BibitemShut {NoStop}%
\bibitem [{\citenamefont {Nordemann}\ \emph {et~al.}(2020)\citenamefont
  {Nordemann}, \citenamefont {Wehrens}, \citenamefont {Tans}, \citenamefont
  {Idema} \emph {et~al.}}]{Idema}%
  \BibitemOpen
  \bibfield  {author} {\bibinfo {author} {\bibfnamefont {G.}~\bibnamefont
  {Nordemann}}, \bibinfo {author} {\bibfnamefont {M.}~\bibnamefont {Wehrens}},
  \bibinfo {author} {\bibfnamefont {S.}~\bibnamefont {Tans}}, \bibinfo {author}
  {\bibfnamefont {T.}~\bibnamefont {Idema}},  \emph {et~al.},\ }\href@noop {}
  {\bibfield  {journal} {\bibinfo  {journal} {arXiv:2003.10509}\ } (\bibinfo
  {year} {2020})}\BibitemShut {NoStop}%
\bibitem [{\citenamefont {Dell’Arciprete}\ \emph {et~al.}(2018)\citenamefont
  {Dell’Arciprete}, \citenamefont {Blow}, \citenamefont {Brown},
  \citenamefont {Farrell}, \citenamefont {Lintuvuori}, \citenamefont {McVey},
  \citenamefont {Marenduzzo},\ and\ \citenamefont {Poon}}]{Blow_Poon}%
  \BibitemOpen
  \bibfield  {author} {\bibinfo {author} {\bibfnamefont {D.}~\bibnamefont
  {Dell’Arciprete}}, \bibinfo {author} {\bibfnamefont {M.}~\bibnamefont
  {Blow}}, \bibinfo {author} {\bibfnamefont {A.}~\bibnamefont {Brown}},
  \bibinfo {author} {\bibfnamefont {F.}~\bibnamefont {Farrell}}, \bibinfo
  {author} {\bibfnamefont {J.~S.}\ \bibnamefont {Lintuvuori}}, \bibinfo
  {author} {\bibfnamefont {A.}~\bibnamefont {McVey}}, \bibinfo {author}
  {\bibfnamefont {D.}~\bibnamefont {Marenduzzo}}, \ and\ \bibinfo {author}
  {\bibfnamefont {W.~C.}\ \bibnamefont {Poon}},\ }\href@noop {} {\bibfield
  {journal} {\bibinfo  {journal} {{Nature Commun.}}\ }\textbf {\bibinfo
  {volume} {9}},\ \bibinfo {pages} {1} (\bibinfo {year} {2018})}\BibitemShut
  {NoStop}%
\end{thebibliography}%

\end{document}